\documentclass[twocolumn,superscriptaddress,floatfix,unsortedaddress,aps,pre,shownopacs,amssymb,amsmath]{revtex4}
\usepackage{graphics}
\usepackage{color}
\usepackage{epsfig}
\usepackage{epstopdf}

\newcommand\ddfrac[2]{\frac{\displaystyle #1}{\displaystyle #2}}

\newcommand{\deltaf}{{\delta\!f}}

\begin{document}

\title{Single-molecule stretching experiments of flexible (wormlike) chain molecules in different ensembles: Theory and a potential application of finite chain length effects to nick-counting in DNA}

\author{Ralf Everaers}
\email{ralf.everaers@ens-lyon.fr}
\affiliation{Universit\'e Lyon, ENS de Lyon, CNRS, Laboratoire de Physique and Centre Blaise Pascal, F-69342 Lyon, France}
\author{Nils B. Becker}
\email{nils.becker@dkfz.de}
\affiliation{German Cancer Research Center, Neuenheimer Feld 580, D-69120 Heidelberg, Germany}
\author{Angelo Rosa}
\email{anrosa@sissa.it}
\affiliation{Scuola Internazionale Superiore di Studi Avanzati (SISSA), Via Bonomea 265, 34136 Trieste, Italy}

\begin{abstract}
We propose a formalism for deriving force-elongation and elongation-force relations for flexible chain molecules from analytical expressions for their radial distribution function, which provides insight into the factors controlling the asymptotic behavior and finite chain length corrections.
In particular, we apply this formalism to our previously developed interpolation formula for the wormlike chain end-to-end distance distribution.
The resulting expression for the asymptotic limit of infinite chain length is of similar quality as the numerical evaluation of Marko's and Siggia's variational theory and considerably more precise than their interpolation formula. 
A comparison to numerical data suggests, that our analytical expressions for the finite-chain length corrections are of similar quality. 
As an application of our results we discuss the possibility of inferring the changing number of nicks in a double-stranded DNA molecule in single-molecule stretching experiments from the accompanying changes in the effective chain length.
\end{abstract}
%


%
\maketitle
%

\section{Introduction}\label{sec:Intro}
\begin{figure}[t]
\includegraphics[width=0.45\textwidth]{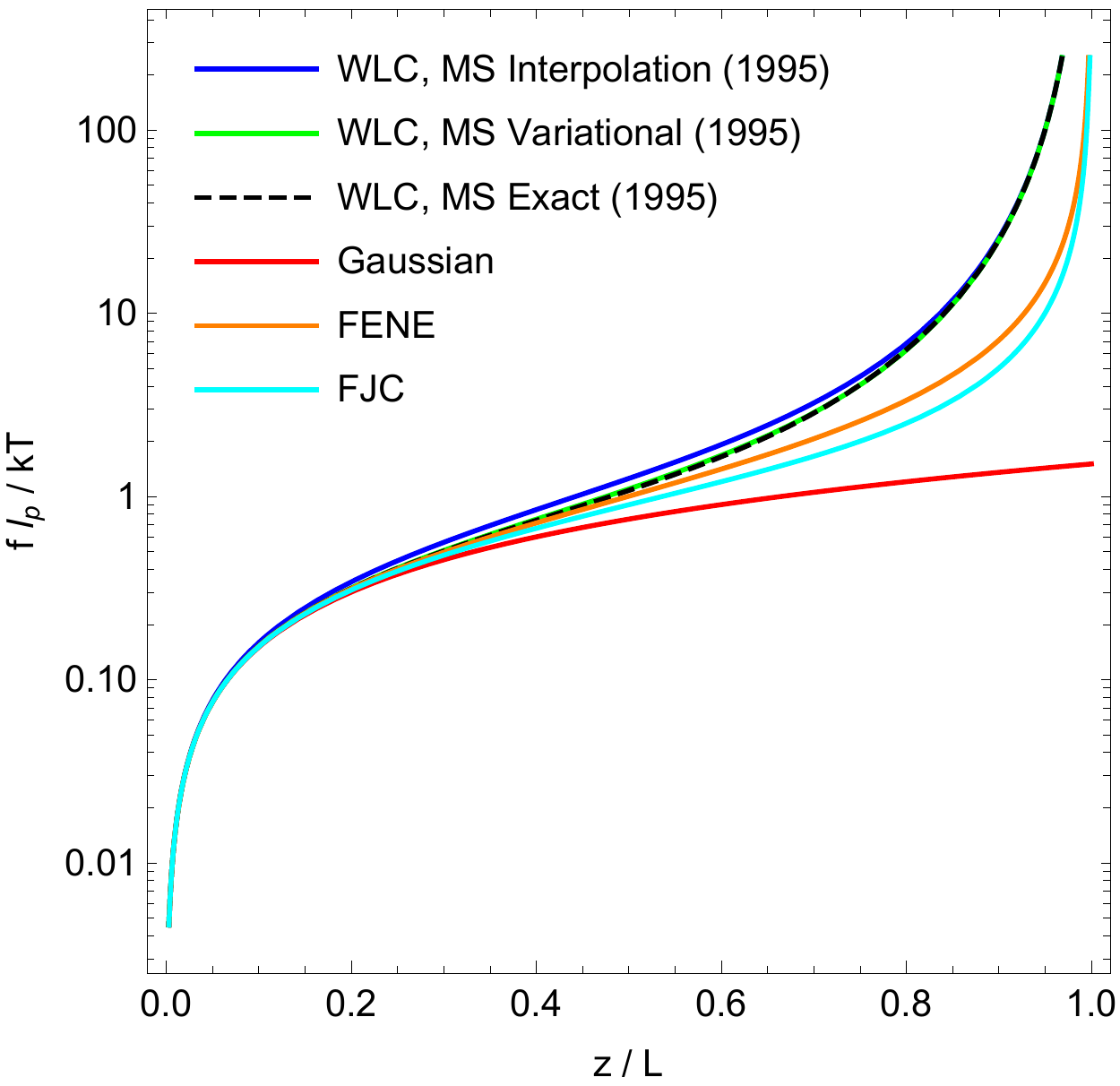}
\caption{
\label{fig:Comparison f(z)}
Comparison of force-elongation curves for a number of popular polymer models discussed in the present article. The abbreviation FJC stands for ``freely-jointed chains''~\cite{DoiEdwards}.
}
\end{figure}

The wormlike chain (WLC)~\cite{KratkyPorod1949} is the standard model for describing the Statistical Physics of semiflexible polymers and is widely used in the context of Biological Physics to describe stiff cytoskeletal filaments like actin or microtubules~\cite{GittesMickeyHowardJCB1993,MorseEntanglSolsSemiFlexPRE1998,KroyFreyPRL1998,EveraersJulicherMaggsPRL1999,MorseDiluteSolsSemiFlexPRE2001,WilhelmFreyStiffPolymerNetworksPRL2003,PampaloniLattanziFreyPNAS2006,HeussingerSchaeferFreyPRE2007,BlundellTerentjev2009,HeussingerSchuellerFreyPRE2010,WenJamneyReview2011,CiccottiPierleoniPolymers2016,MengTerentjevReview2017}.
The present article is primarily motivated by the application of the WLC~\cite{BustamanteScience1994,MarkoSiggia1995,RiefPRL1998}  
to single-molecule experiments, where double-helical DNA~\cite{SmithFinziBustamante1992,BustamanteScience1994,Vologodskii1994,BouchiatCroquetteBJ1999,StrickBensimonReview2002,MoffittBustamanteReview2008},
proteins~\cite{RiefTitinScience1997} or polysaccharides~\cite{RiefPolysaccharidesScience1997} are stretched by an external force.
At least in the case of DNA~\cite{BustamanteScience1994,MarkoSiggia1995}, experiments  are significantly better described by the WLC than by other polymer models (Fig.~\ref{fig:Comparison f(z)}).
Here, we present a systematic derivation of the elastic response of stretched chain molecules in the constant-force and constant-elongation ensembles starting from a given expression for their radial distribution function. Besides recovering the known asymptotic behavior of WLC with excellent precision, 
we obtain high-quality expressions for finite chain length effects in the two ensembles. 
We argue that their knowledge might allow to detect (changes in) the number of single-strand breaks in single-molecule stretching experiments of double-helical DNA.

The paper is organized as follows:
In Section~\ref{sec:Background}, we briefly summarize the main features of the WLC model and the main points of the classical work by Marko and Siggia~\cite{MarkoSiggia1995}.
Section~\ref{sec:Theory} is devoted to a systematic derivation of the elongation-force and force-elongation relations of long (wormlike) polymers from a given expression for the chain end-to-end distance distribution. In particular, we obtain analytical expressions for and gain insight into the finite chain length corrections in the two ensembles, where chains are held at constant force and constant elongation respectively.
As a first validation, we consider in Secs.~\ref{sec:Gaussian Chains} and \ref{sec:FENE} the exactly solvable cases of Gaussian and finite-extensible nonlinear-elastic (FENE)-springs.
In a second step, we apply the formalism to two approximate expressions for the radial distribution function of WLC.
Bhattacharjee, Thirumalai and Bryngelson (BTB)~\cite{BTBpdf-1997} derived a suitable expression using the variational theory of Ha and Thirumalai~\cite{Ha1995,Ha1997}.
As an alternative, we (BRE)~\cite{BeckerRosaEveraers2010} proposed an interpolation between exact results for all relevant limiting cases of the WLC model ranging from short (stiff) to long (flexible) chains and including looped and fully stretched configurations.
In Sections~\ref{sec:BTB} and \ref{sec:BRE} we derive  for the corresponding ``BTB''- and ``BRE''-springs analytical expressions for the asymptotic force-elongation relation and the first-order corrections in both ensembles.
In the discussion in Sec.~\ref{sec:WLC} we compare these expressions to analytical, numerical and simulation results for WLC. In addition, we discuss the elastic response of nicked WLC composed of several freely jointed wormlike segments. In this case, finite chain length effects turn out to be controlled by the average segment length. In particular, we show that under suitable conditions single-molecule stretching experiments of DNA should be able to detect enzyme-induced changes in the number of single-strand breaks.
We briefly conclude in Sec.~\ref{sec:Conclusions}.
The Appendix summarises the simulation and data analysis methods we have used to obtain numerical reference data for FENE-, BTB- and BRE-springs as well as WLC.

%

\section{Background\label{sec:Background}}

\subsection{The model \label{sec:Model}}
The WLC is defined via a Hamiltonian
\begin{equation}\label{eq:WLC Hamiltonian}
\frac{\mathcal{H}_{\rm WLC} }{k_BT}= \frac12 l_p \int_0^L \left(\frac{\partial^2}{\partial s^2} \vec r(s) \right)^2 ds
\end{equation}
for incompressible space-curves of contour length $L$ with bending rigidity $l_p \, k_BT$, where $l_p$ is the persistence length and $k_BT$ the thermal energy.
Despite its simple appearance, the incompressibility constraint, $\left| \frac{\partial}{\partial s} \vec r(s) \right| \equiv1$, renders the model non-trivial so solve.
Notable exceptions~\cite{NagaiPolymers1973} are the even moments $\langle r^{2k}(L) \rangle$ of the end-to-end distance $r$
and,
in particular ($k=1$), the 
mean-square end-to-end distance~\cite{KratkyPorod1949} given by the formula 
\begin{eqnarray}\label{eq:WLC R2}
\langle r^2(L) \rangle &\equiv& \langle \left| \vec r(L) - \vec r(0) \right|^2 \rangle\nonumber\\
& = & 2 \, l_p^2 \, \left( \frac{L}{l_p} + e^{-L/l_p} - 1 \right) \, .
\end{eqnarray}
This expression shows a crossover from rigid rod behaviour, $\langle r^2(L) \rangle = L^2$, to random walk behavior, $\langle r^2(L) \rangle = 2 l_p L$, for contour lengths, $L$,  around the persistence length, $l_p$.

Nature offers examples of polymers in a wide range of ratios $L/l_p$.
For instance, cytoskeletal filaments like microtubules~\cite{PampaloniLattanziFreyPNAS2006} typically have $L\le l_p$.
In this work we focus on chains, which are much longer than their persistence length, $L\gg l_p$ and $N_p \equiv L/l_p \gg1$.
This is, for example, the case in DNA stretching experiments~\cite{MarkoSiggia1995}.

We are interested in two related mechanical problems, the force-elongation and the elongation-force relation of freely rotating WLC.
The former specifies the expectation value of the force, $\langle \vec f(z) \rangle= \langle f(z) \rangle \, \vec e_z $, required to constrain the projected elongation of a WLC to a constant value $z\equiv z(L)-z(0) = \left(\vec r(L)-\vec r(0) \right)\cdot \vec e_z$.
The latter denotes the average elongation, $\langle z(f) \rangle$, of a WLC in the direction of a constant force, $\vec f = f \vec e_z$, separating its ends.
That is, the average $\langle z(f) \rangle$ is taken with respect to the forced Hamiltonian
\begin{equation}\label{eq:WLC under force Hamiltonian}
\mathcal{H} = \mathcal{H}_{\rm WLC} - f z \, .
\end{equation}

\subsection{DNA stretching \label{sec:MarkoSiggia}}
In their seminal analysis~\cite{MarkoSiggia1995}, Marko and Siggia discussed {\it inter alia}:
1)
the asymptotic behaviour of the force in the limit of strong stretching
\begin{equation}\label{eq:z(f) WLC large f}
\lim_{z\rightarrow L} \frac{f}{k_BT/l_p} = \frac1{4(1-z/L)^2} \, ;
\end{equation}
2) an analytic expression (blue in Fig.~1),
\begin{equation}
\label{eq:MarkoSiggia f(z)}
\frac{f}{k_BT/l_p} = \frac zL + \frac1{4(1-z/L)^2} -\frac14 \, ,
\end{equation}
interpolating from Eq.~(\ref{eq:z(f) WLC large f}) to the opposite (random walk) limit of weak stretching
\begin{equation}\label{eq:z(f) WLC small f}
\lim_{z\rightarrow 0} \frac{f}{k_BT/l_p} = \frac32  \frac zL \, ;
\end{equation}
3) a more precise variational calculation of the stretching force (green in Fig.~1); 
4) how to obtain the exact force-elongation relation with sufficient precision by numerically diagonalising a $100\times100$ matrix (dashed black in Fig.~1). 
	They noted that 3) and 4) were necessary because of the high quality of the experimental data.

Marko and Siggia worked in the constant-tension ensemble, since they were motivated~\cite{BustamanteScience1994} by the experiments of Smith {\it et al.}~\cite{SmithFinziBustamante1992}, who attached one end of phage-$\lambda$ DNA to a glass slide and the other to a magnetic bead on which they could exert a force.
The complementary constant-elongation 
ensemble can be explored in atomic force microscope experiments~\cite{RiefGaubNatStrBiol1999,Clausen-SchaumannGaubBJ2000}, where the mobile end of the DNA molecule is attached to a cantilever, which probes the force needed to maintain an imposed constant displacement.
The results of pulling experiments in the two ensembles are not expected to be equivalent for chains of finite contour lengths; 
that is, the force-elongation relation in the constant-force ensemble is not the inverse function of the elongation-force relation in the constant-elongation ensemble~\cite{NeumannBJ2003,HolmPRE2009,Ivanov2012}. 
Below we will consider both situations in turn (see Secs.~\ref{sec:ForceElongation} and~\ref{sec:ElongationForce}).

\section{Theory\label{sec:Theory}}

In the present work we infer the elastic properties of polymers from their end-to-end distance distribution,
\begin{equation}
Q(r) \equiv \frac1{4\pi r^2} \langle \delta\left(\left| \vec r(L) - \vec r(0) \right| -r  \right) \rangle \  .
\end{equation}
We first consider experiments performed in the constant-elongation ensemble. The second part of this section deals with the constant-force ensemble.

\subsection{Force-elongation relations from end-to-end distance distributions}\label{sec:ForceElongation}
For chains whose free ends are constrained to a particular $z$-plane, the partition function is
\begin{eqnarray}\label{eq:WLC Partition function constrained zL}
\mathcal{Z}(z)
&\propto& \int d\vec r' \,  Q(r') \delta \left( z' - z \right) \nonumber\\
&\propto& \int_0^{\sqrt{L^2-z^2}} d\rho\,  \rho\, Q\left(\sqrt{\rho^2+z^2}\right) . 
\end{eqnarray}
We define the potential of mean force as
\begin{equation}
\mathcal{F}(z) = - k_BT \log{\mathcal Z}(z)
\label{eq:Z(z)final}
\end{equation}
so that the mean required constraining force is
\begin{equation}\label{eq:f(z)}
\langle f(z) \rangle = \frac d{dz} \mathcal{F}(z) = -k_BT \,\frac{{\mathcal Z}'(z)}{{\mathcal Z}(z)}\ .
\end{equation}
Without loss of generality, we consider chain ends constrained at $z>0$.
As a consequence of our force convention, the constraining forces are also positive, $f>0$.

End-to-end distance distributions, $Q(r;L,l_p)=Q(r/L,l_p/L)$, and partition functions, $\mathcal{Z}(z;L,l_p)=\mathcal{Z}(z/L,l_p/L)$, can be written as a function of two dimensionless variables:
the chain elongation, $\zeta=z/L$, in units of the maximal elongation and
the inverse of the chain length in units of the bending persistence length, $\kappa = 1/N_p = l_p/L$.
To proceed, we perform an analogous switch from the extensive free energy $ \mathcal{F}$ to an intensive free energy $\mathcal{F}_p$ per persistence length:
\begin{eqnarray}
\mathcal{F}(z;L,l_p) &\equiv& N_p  \mathcal{F}_p(\zeta,\kappa)\\
\mathcal{F}_p(\zeta,\kappa) &=& - k_BT \log\left( \mathcal{Z}_p(\zeta,\kappa)\right)\\
\mathcal{Z}_p(\zeta,\kappa) &\equiv&  \mathcal{Z}(z;L,l_p)^{\kappa} \ .
\end{eqnarray}
Using this notation,
constraining forces can be computed as
\begin{eqnarray}
\langle f(z;L,l_p) \rangle
&=& \frac d{dz} \mathcal{F}(z;L,l_p)\nonumber\\
&=& N_p \frac d{dz} \mathcal{F}_p(\zeta=z/L,\kappa)\nonumber\\
&=& \frac1{l_p} \mathcal{F}_p^{(1,0)}(\zeta,\kappa)\nonumber\\
&=& -\frac{k_BT}{l_p} \frac{\mathcal{Z}_p^{(1,0)}(\zeta,\kappa)}{\mathcal{Z}_p(\zeta,\kappa)},
\end{eqnarray}
where we have introduced the notation $X^{(i,j)}\equiv\partial_\zeta^i\partial_\kappa^j X$ 
for quantities $X=X(\zeta,\kappa)$.
Constraining forces can be directly expressed in the natural units of force, $k_BT/l_p$:
\begin{equation}\label{eq:phi}
\langle \phi(\zeta,\kappa) \rangle \equiv \frac{\langle f(z;L,l_p) \rangle}{k_BT/l_p}
= - \frac{\mathcal{Z}_p^{(1,0)}(\zeta,\kappa)}{\mathcal{Z}_p(\zeta,\kappa)}
= - \kappa \frac{\mathcal{Z}^{(1,0)}(\zeta,\kappa)}{\mathcal{Z}(\zeta,\kappa)}.
\end{equation}

\subsubsection{Asymptotic behavior and finite-size corrections}
For chains, which are much longer than their persistence length, $L\gg l_p$ and $N_p\gg1$, the parameter $\kappa \ll1$ can serve as a convenient expansion parameter for identifying the behavior close to the thermodynamic limit of infinitely long chains:
\begin{eqnarray}
\mathcal{F}_p(\zeta,\kappa) &=& \mathcal{F}_p(\zeta) + \delta\mathcal{F}_p(\zeta,\kappa)\\
\delta\mathcal{F}_p(\zeta,\kappa)
&=& \sum_{n=1}^\infty  \left. \frac{\kappa^n}{n!} \frac{d^n \mathcal{F}_p(\zeta,\kappa)}{d\kappa^n}\right|_{\kappa=0}\\
&\equiv& \sum_{n=1}^\infty  \frac{\kappa^n}{n!} \mathcal{F}_p^{(0,n)}(\zeta,0)\ .
\end{eqnarray}
Retaining the leading term
\begin{eqnarray}
\mathcal{F}_p(\zeta) &=& -k_BT \log\left( \mathcal{Z}_{p}(\zeta,0) \right)
\end{eqnarray}
and corrections to first order in $\kappa$,
\begin{eqnarray}
\delta\mathcal{F}_p(\zeta,\kappa)
&\approx& \kappa\, \mathcal{F}_p^{(0,1)}(\zeta,0) \\
 &=&  - \kappa\ k_BT \ddfrac{\mathcal{Z}_{p}^{(0,1)}(\zeta,0)}{ \mathcal{Z}_{p}(\zeta,0)}
\end{eqnarray}
the corresponding force-elongation relation reads
\begin{eqnarray}
\langle \phi(\zeta,\kappa) \rangle &=&  \phi(\zeta) + \delta \phi(\zeta,\kappa)\\
\phi(\zeta) &=& - \frac{\mathcal{Z}_{p}^{(1,0)}(\zeta,0)}{\mathcal{Z}_{p}(\zeta,0)}
\label{eq:f of Zp}\\
\delta \phi(\zeta,\kappa) &\approx&
  \kappa \left( -  \ddfrac{\mathcal{Z}_{p}^{(1,1)}(\zeta,0)}{ \mathcal{Z}_{p}(\zeta,0)} \right.
\label{eq:delta f of Zp}\\
&&  \left. +   \ddfrac{\mathcal{Z}_{p}^{(0,1)}(\zeta,0)}{ \mathcal{Z}_{p}(\zeta,0)}  \ddfrac{\mathcal{Z}_{p}^{(1,0)}(\zeta,0)}{ \mathcal{Z}_{p}(\zeta,0)} \right)\ .
\nonumber
\end{eqnarray}
In particular, corrections are of the order $\deltaf \sim \kappa \frac{k_BT}{l_p} = \frac1{N_p}  \frac{k_BT}{l_p} =  \frac{k_BT}{L}$.

\subsubsection{Approximations \label{sec:ForceElongation Approximations}}
Since we are not always able to carry out the integrations in Eq.~(\ref{eq:WLC Partition function constrained zL}), we develop an approximation scheme valid for long chains, $L\gg l_p$ and $N_p\gg1$, where $Q(r)$ is a monotonically decreasing function of distance.

Neglecting fluctuations, we may restrict the partition function to conformations with the minimal end-to-end distance, $r=z$, at the considered elongation in $z$-direction.
Denoting the partition function for chains with $z$-aligned end-to-end vectors by $\mathcal Z_{(\cdot)}$, we approximate
\begin{eqnarray}\label{eq:Zdot}
\mathcal{Z}(\zeta,\kappa) \approx \mathcal{Z}_{(\cdot)}(\zeta,\kappa) \propto Q(\zeta,\kappa)\ .
\end{eqnarray}

With $\mathcal{Z}_{p,(\cdot)}(\zeta,\kappa) \propto Q_p(\zeta,\kappa)$ the corresponding zeroth and first order contributions to the restoring force can be directly read off from Eqs.~(\ref{eq:f of Zp}) and (\ref{eq:delta f of Zp}):
\begin{eqnarray}
\phi_{(\cdot)}(\zeta,\kappa) &\equiv&  \phi_{(\cdot)}(\zeta) + \delta\phi_{(\cdot)}(\zeta,\kappa)\\
\phi_{(\cdot)}(\zeta) &=& - \frac{Q_{p}^{(1,0)}(\zeta,0)}{Q_{p}(\zeta,0)}
\label{eq:fdot of Qp}\\
\delta\phi_{(\cdot)}(\zeta,\kappa)&\approx&
  \kappa \left(- \ddfrac{Q_{p}^{(1,1)}(\zeta,0)}{ Q_{p}(\zeta,0)} \right.
\label{eq:delta fdot of Qp}\\
&&  \left. +   \ddfrac{Q_{p}^{(0,1)}(\zeta,0)}{Q_{p}(\zeta,0)}  \ddfrac{Q_{p}^{(1,0)}(\zeta,0)}{ Q_{p}(\zeta,0)} \right).
\nonumber
\end{eqnarray}

In a second step, we can approximate the integration over the transverse degrees of freedom by expanding
$-k_BT\log\left(Q\left(\sqrt{\rho^2+\zeta^2},\kappa\right)\right) \approx -k_BT \log\left(Q\left(\zeta,\kappa\right)\right) + \frac12 k_{(\perp)}(\zeta,\kappa) \rho^2$,
the linear contribution being absent because the displacement in $\rho$-direction is perpendicular to the elongation in $z$-direction.
The coefficient of the second order term,
\begin{eqnarray}
k_{(\perp)}(\zeta,\kappa)
&=& -k_BT \left.\frac{\partial^2}{\partial \rho^2} \log\left(Q\bigl(\sqrt{\rho^2+\zeta^2},\kappa\bigr)\right)\right|_{\rho=0}\nonumber\\
	&=& -\frac{k_BT}{\zeta} \frac{Q^{(1,0)}(\zeta,\kappa)}{Q(\zeta,\kappa)} = \frac{f_{(\cdot)}(\zeta,\kappa)}{\zeta}
\label{eq:kperp}
\end{eqnarray}
is the effective stiffness at the minimal elongation $r=z$ of chains constrained to a particular $z$-plane.
Note that independently of chain length we are dealing with a single degree of freedom and that we are expanding (an approximation of) the extensive partition function, $Q$, and not $Q_p$.
We remark that Eq.~(\ref{eq:kperp}) is identical to the expression derived by Strick {\it et al.}~\cite{Strick1998} and used to measure the force $f$ exerted on DNA molecules pulled by magnetic beads. 

Extending the limits of the $\rho$-integration to infinity and carrying out the Gaussian integral,
\begin{eqnarray}\label{eq:ZdotZperp}
\mathcal{Z}(\zeta,\kappa) &\approx& \mathcal{Z}_{(\cdot)}(\zeta,\kappa) \mathcal{Z}_{(\perp)}(\zeta,\kappa)\\
\mathcal{Z}_{(\perp)}(\zeta,\kappa)
&\propto&  \frac{1}{k_{(\perp)}(\zeta,\kappa) } \propto  \frac{\zeta} {\phi_{(\cdot)}(\zeta,\kappa)}\ .
\end{eqnarray}
The change of the transverse fluctuations upon stretching makes an additive contribution to the restoring force,
\begin{eqnarray}
\langle f(\zeta,\kappa) \rangle &\approx& f_{(\cdot)}(\zeta,\kappa) + f_{(\perp)}(\zeta,\kappa)\ .
\end{eqnarray}
With $\phi_{(\perp)}(\zeta,\kappa) =  - \kappa \frac{\mathcal{Z}_{(\perp)}^{(1,0)}(\zeta,\kappa)}{\mathcal{Z}_{(\perp)}(\zeta,\kappa)}$, Eq.~(\ref{eq:phi}), this contribution vanishes asymptotically,
\begin{eqnarray}
\phi_{(\perp)}(\zeta,\kappa) &\equiv&  \phi_{(\perp)}(\zeta) + \delta\phi_{(\perp)}(\zeta,\kappa)\\
\phi_{(\perp)}(\zeta) &=& 0
\label{eq:fperp of Qp}
\end{eqnarray}
so that the asymptotic force-elongation relation is given by  $\phi_{(\cdot)}(\zeta)$ and Eq.~(\ref{eq:fdot of Qp}).
Furthermore, we may neglect corrections to $\phi_{(\cdot)}(\zeta)$ when evaluating
\begin{eqnarray}
\delta\phi_{(\perp)}(\zeta,\kappa)
&\approx& \kappa  \left(  \frac{\phi_{(\cdot)}^{(1,0)}(\zeta,\kappa)}{\phi_{(\cdot)}(\zeta,\kappa)} - \frac 1\zeta  \right)\nonumber\\
&\approx& \kappa  \left( \frac{\phi_{(\cdot)}^{(1,0)}(\zeta)}{\phi_{(\cdot)}(\zeta)} - \frac 1\zeta  \right)
\label{eq:delta phi perp}
\end{eqnarray}
to first order in $\kappa$.

\subsection{Elongation-force relations from end-to-end distance distributions}\label{sec:ElongationForce}
For chains stretched in $z$-direction
\begin{eqnarray}
\mathcal{Z}(f)
&\propto& \int d\vec r \,  Q(r)\exp\left(\frac{f\, z}{k_BT}\right)
\nonumber\\
&=& \int_{-L}^L dz \,  \mathcal{Z}(z)  \exp\left(\frac{f\, z}{k_BT}\right) . \label{eq:Partition function under force}
\end{eqnarray}
We define the potential of mean elongation
\begin{equation}
	\mathcal{G}(f) = - k_BT \log\left(\mathcal{Z}(f) \right), \label{eq:DefineGf}
\end{equation}
so that
\begin{eqnarray}
\langle z(f) \rangle
   &=& \frac{\int_{-L}^L dz \,  \mathcal{Z}(z)\,z \exp\left(\frac{f\, z}{k_BT}\right)}{\int_{-L}^L dz \,  \mathcal{Z}(z)\exp\left(\frac{f\, z}{k_BT}\right) }
   \label{eq:z(f)}\\
   &=& -\frac{d}{d f} \mathcal{G}(f)
   \label{eq:z(f) from G}
\end{eqnarray}
Again, we are not always able to carry out the integrations in Eqs.~(\ref{eq:Partition function under force}) and (\ref{eq:z(f)}), forcing us to generalize the above approximation scheme to fluctuations in $z$-direction.

\subsubsection{Finite-size corrections to the inverse of the asymptotic force-elongation relation}
Given a force-elongation relation, $\langle f(z) \rangle$, approximate elongation-force relations, $\langle z(f) \rangle$, can be obtained from Eqs.~(\ref{eq:Partition function under force}) to (\ref{eq:z(f) from G}) with the help of Laplace's method.
Expanding the logarithmic integrand of ${\mathcal Z}(f)$ around $0\le z^\ast \le L$ and noting that
$f^{(n-1)}(z^\ast)=\mathcal{F}^{(n)}(z^\ast)$:
\begin{eqnarray}\label{eq:F expansion}
-\frac{\mathcal{F}(z)}{k_BT} +\frac{f z}{k_BT}
&= &-\frac{\mathcal{F}(z^\ast)}{k_BT} +\frac{f z^\ast}{k_BT} \\
&& +\left( -\frac{f(z^\ast)}{k_BT} + \frac{f}{k_BT} \right)\, \left(z-z^\ast\right)\nonumber\\
&& -\frac12 \frac{f'(z^\ast)}{k_BT} \left(z-z^\ast\right)^2\ \nonumber\\
&& -\frac16 \frac{f''(z^\ast)}{k_BT} \left(z-z^\ast\right)^3\ \nonumber\\
&& -\sum_{n=4}^\infty \frac1{n!}\frac{f^{(n-1)}(z^\ast)}{k_BT} \left(z-z^\ast\right)^n \, .\nonumber
\end{eqnarray}
The integrand develops a maximum at $z^\ast$, if the external force, $f = f(z^\ast) = {\mathcal F}'(z^\ast)$, is equal to the average force (see Eq.~(\ref{eq:f(z)})) required to constrain the elongation to $z^\ast$.
The second order term describes the longitudinal stiffness at this elongation with an effective spring constant of $k_{(||)}(z^\ast) = \mathcal{F}''(z^\ast)/k_BT= f'(z^\ast)/k_BT$.
The third order term gives rise to anisotropic fluctuations around $z^\ast$.

\begin{figure}[t]
\includegraphics[width=0.34\textwidth,angle=270]{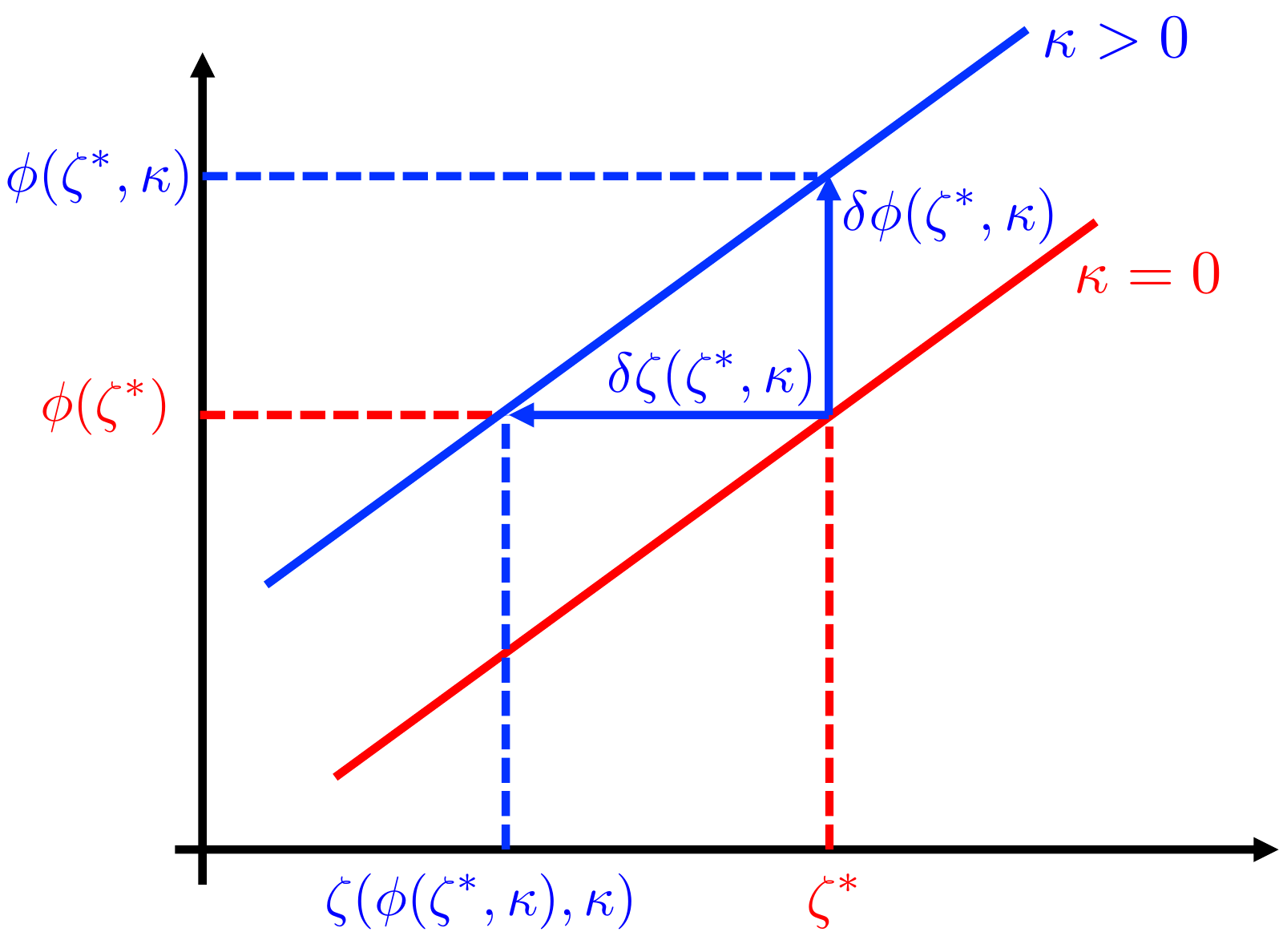}
\caption{
``Geometric'' derivation of Eqs.~(\ref{eq:InvertPhiXi-FirstOrderCorrections-1}) and~(\ref{eq:InvertPhiXi-FirstOrderCorrections-2}):
conversion of the  first-order correction, $\delta \phi(\zeta^\ast,\kappa)$, to the force-elongation relation, to the first order correction, $\delta\zeta(\zeta^\ast,\kappa)$, of the inverse force-elongation relation.
\label{fig:InvertPhiXi-FirstOrderCorrections}
}
\end{figure}

To a first approximation, which becomes exact in the asymptotic limit, one can neglect all terms beyond the linear order, $n=1$:
\begin{eqnarray}
	\mathcal{Z}(f(z^\ast)) &\approx& \mathcal{Z}(z^\ast) \exp\left( \frac{f(z^\ast)z^\ast}{k_BT}  \right)\\
\mathcal{G}(f) &\approx& \mathcal{F}(z^\ast) - f z^\ast \, .
\end{eqnarray}
While this trivially equates the elongation-force relation with the inverted force-elongation relation, {\it i.e.} in natural units
\begin{eqnarray}
\zeta\left(\phi(\zeta^\ast,\kappa),\kappa  \right) \equiv \zeta^\ast \, ,
\end{eqnarray}
we are still left with two problems.
First, we are not necessarily able to invert a general, non-linear force-elongation relation in closed form:
in such cases, we can still provide a parametric representation
of the elongation-force curve.
In particular, in the asymptotic limit of $\kappa=0$ we can plot $\{\phi(\zeta^\ast),  \zeta(\phi(\zeta^\ast)) \equiv \zeta^\ast \}$ for $0\le \zeta^\ast \le 1$.
Second, we need to convert our corrections, $\delta\phi(\zeta,\kappa)$ to the asymptotic force-elongation relation into corresponding corrections $\delta \zeta(\phi,\kappa)$ to the asymptotic elongation-force relation.
As illustrated in Fig.~\ref{fig:InvertPhiXi-FirstOrderCorrections}, we may write to first order in $\kappa$:
\begin{eqnarray}
 \zeta\left( \phi(\zeta^\ast,\kappa), \kappa \right)  &=& \zeta^\ast + \delta\zeta(\zeta^\ast,\kappa) \, , \label{eq:InvertPhiXi-FirstOrderCorrections-1} \\
\delta\zeta(\zeta^\ast,\kappa) &\approx& - \frac{\delta \phi(\zeta^\ast,\kappa)}{\phi'(\zeta^\ast)} \, . \label{eq:InvertPhiXi-FirstOrderCorrections-2} 
\end{eqnarray}
In particular,
\begin{eqnarray}
\delta\zeta_{(\cdot)}(\zeta^\ast,\kappa) &\approx& - \frac{\delta \phi_{(\cdot)}(\zeta^\ast,\kappa)}{\phi'(\zeta)} \, , \label{eq:delta zeta dot}\\
\delta\zeta_{(\perp)}(\zeta^\ast,\kappa) &\approx& - \frac{\delta \phi_{(\perp)}(\zeta^\ast,\kappa)}{\phi'(\zeta)} \, .
\label{eq:delta zeta perp}
\end{eqnarray}

\subsubsection{Additional finite-size corrections}
Higher order terms in Eq.~(\ref{eq:F expansion}) describe additional corrections, $\delta\zeta_{(||)}(\zeta^\ast,\kappa)$, induced by longitudinal fluctuations:
\begin{eqnarray}\label{eq:all corrections to z(f)}
\lefteqn{\langle \zeta\left( \phi(\zeta^\ast,\kappa)  \right) \rangle =}\\
&& \zeta^\ast + \delta\zeta_{(\cdot)}(\zeta^\ast,\kappa)+ \delta\zeta_{(\perp)}(\zeta^\ast,\kappa)+ \delta\zeta_{(||)}(\zeta^\ast,\kappa) \, . \nonumber 
\end{eqnarray}
To a second approximation, we retain the second order term in Eq.~(\ref{eq:F expansion}) and expand the exponential with the third order term to first order
\begin{eqnarray}
\lefteqn{\mathcal{Z}(z) \exp\left( \frac{f(z^\ast)z}{k_BT}  \right) \approx }\nonumber\\
&& \mathcal{Z}(z^\ast) \exp\left( \frac{f(z^\ast)z^\ast}{k_BT}  \right) \times\\
&& \exp\left(-\frac12 \frac{f'(z^\ast)}{k_BT} \delta z^2 \right)  \left(1 -\frac16 \frac{f''(z^\ast)}{k_BT} \delta z^3\right)\nonumber
\end{eqnarray}
when evaluating the integrals in Eqs.~(\ref{eq:Partition function under force}) and (\ref{eq:z(f)}).
Extending the integration range to infinity and noting that Gaussian integrals for odd powers of $\delta z$ vanish due to symmetry reasons, the partition function for longitudinal fluctuations is given by
\begin{eqnarray}
 \mathcal{Z}_{(||)}(f(z^\ast))  = \sqrt{\frac{2\pi k_BT}{f'(z^\ast)}}\ .
\end{eqnarray}
while
\begin{eqnarray}\label{eq:Approxiate z(f)}
\label{eq:delta delta z}
\delta z_{(||)}(z^\ast) &=& -\frac{k_BT}2 \frac{f''(z^\ast)}{\left(f'(z^\ast)\right)^2} \ .
\end{eqnarray}
Note that the latter result can also be obtained by differentiating the corrected force-dependent free energy,
\begin{eqnarray}
\mathcal{G}(f)
&=& \mathcal{F}(z^\ast) - f z^\ast+\frac{k_BT}2 \log\left( \frac{f'(z^\ast)}{k_BT} \right)\ .
\end{eqnarray}
with respect to the applied force, Eq.~(\ref{eq:z(f) from G}).
Rewriting in terms of dimensionless variables and to first order in $\kappa$,
\begin{eqnarray}
\delta \zeta_{(||)}(\zeta^\ast,\kappa) =   \kappa \frac{\phi''(\zeta^\ast)}{2\left(\phi'(\zeta^\ast)\right)^2}\ ,
\label{eq:delta zeta par}
\end{eqnarray}
we see that this effect is of comparable magnitude to the other corrections in Eq.~(\ref{eq:all corrections to z(f)}).

To summarize, we have identified three main contributions entering into the first-order corrections to the asymptotic chain behavior:
(1)
from chain conformations whose end-to-end vectors is aligned along the $z$-direction (symbol $(\cdot)$);
(2)
from chain conformations whose end-to-end vectors make ``transverse'' fluctuations, orthogonal to the prescribed $z$-direction (symbol $(\perp)$); 
(3)
from chain conformations fluctuating along the $z$-direction, {\it i.e.} longitudinal fluctuations which are allowed {\it only} in the constant force ensemble (symbol $(||)$).

\section{Stretching Gaussian springs \label{sec:Gaussian Chains}}
As a first sanity check,  we apply the above formalism to the ubiquituous~\cite{DoiEdwards} Gaussian chain model of polymer physics, which describes the conformations of long, $L\gg l_p$, non-interacting or ideal chains, whose radial distribution function follows a Gaussian distribution as long as $r\ll L$:
\begin{equation}\label{eq:QG}
Q(r)  \propto \exp\left( -\frac{3 r^2}{4 l_p L}  \right) = \exp\left( -\frac34 N_p \left(\frac rL\right)^2 \right) \, .
\end{equation}

\subsection{Exact solution}
Adopting the Gaussian chain model for arbitrary distances, different spatial dimensions remain uncoupled.
As a consequence, $\mathcal{Z}(z) \propto \exp\left( -\frac34 N_p \left(\frac zL\right)^2 \right)$, so that the force-elongation relation,
\begin{eqnarray}\label{eq:QG f(z)}
\frac{\langle f(z) \rangle}{k_BT/l_p}  =   \frac32 \frac{z}{L} \ ,
\end{eqnarray}
is straightforward to calculate exactly.
Similarly, with
$\mathcal{Z}(f) \propto  \exp\left(\frac13 N_p \left(\frac{f l_p}{k_BT} \right)^2\right)$,
one obtains
\begin{eqnarray}\label{eq:QG z(f)}
\frac{\langle z(f) \rangle}{L} =   \frac23 \frac{f}{k_BT/l_p}\ .
\end{eqnarray}
for the elongation-force relation.

\subsection{Asymptotic behavior}
As there are no finite-size corrections to the force-elongation and elongation-force relations of Gaussian chains, Eqs.~(\ref{eq:QG f(z)}) and (\ref{eq:QG z(f)}), the two relations are each other's inverse.

Following the analysis in Sec.~\ref{sec:ForceElongation Approximations} we should be able to derive the same result from the asymptotic partition function per persistence length for chains extended to the minimal end-to-end distance, $r=z$. With
\begin{eqnarray}
\mathcal{Z}_{p,(\cdot)}(z/L)
&\propto&  \exp\left( -\frac34 \left(\frac zL\right)^2 \right)\ ,
\end{eqnarray}
it is straightforward to see that Eq.~(\ref{eq:fdot of Qp}) yields indeed the correct result:
\begin{eqnarray}
\frac{ f_{(\cdot)}(z/L)}{k_BT/l_p} &=&  \frac32 \frac{z}{L}\ .
\end{eqnarray}

\subsection{Finite chain length corrections}
Do we understand the absence of corrections?
It turns out that $\mathcal{Z}_{p,(\cdot)}(z/L,\kappa)  = \mathcal{Z}_{p,(\cdot)}(z/L)$ independently of chain length.
As there are no finite size corrections to  the dominant free energy contribution, $\mathcal{F}_{p,(\cdot)}(z/L)$, there are also no corresponding corrections, Eq.~(\ref{eq:delta fdot of Qp}),  to the elastic response
\begin{eqnarray}
\frac{ \deltaf_{(\cdot)}(z/L)}{k_BT/l_p} &=&  0\ .
\end{eqnarray}
Similarly,  there is neither a finite-size correction, Eq.~(\ref{eq:delta phi perp}), to the force-elongation curve due to transverse fluctuations,
\begin{eqnarray}
\frac{ \deltaf_{(\perp)}(z/L)}{k_BT/l_p} &=&  0\ ,
\end{eqnarray}
nor a correction to the elongation-force relation, Eq.~(\ref{eq:delta zeta par}), 
due to longitudinal fluctuations,
\begin{eqnarray}
\frac{ \delta z_{(||)}(z/L)}{L} &=&  0\ ,
\end{eqnarray}
since the asymptotic force-elongtion, $f(z)=k z$, is harmonic.

\begin{figure*}
\includegraphics[width=\textwidth]{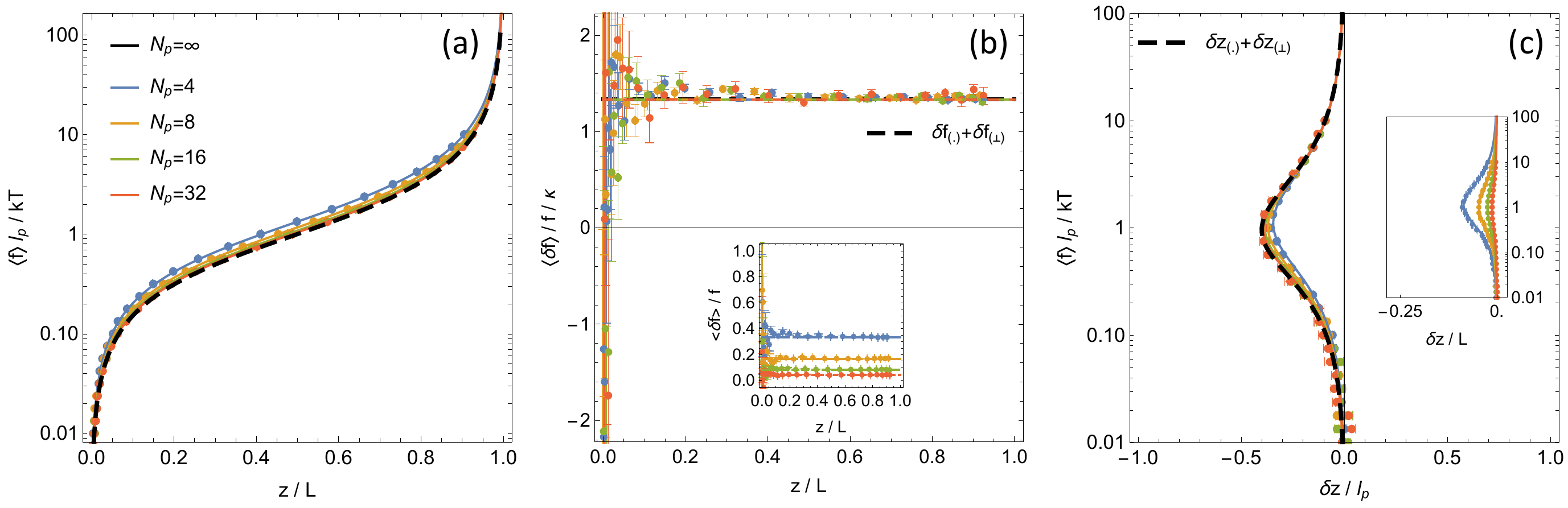}
\caption{
\label{fig:FENE f(z)}
FENE-springs in the constant-elongation ensemble:
(a)
force-elongation relations, 
(b)
finite-size corrections ($\langle \delta f \rangle \equiv \langle f(z, \kappa)\rangle - \langle f(z, \kappa=0)\rangle$) to the force-elongation relation, 
(c)
finite-size corrections ($\delta z \equiv \langle f(z, \kappa) \rangle^{-1} - \langle f(z, \kappa=0) \rangle^{-1}$) to the inverted force-elongation relation. 
The insets show finite-size corrections in the units of the force-elongation relation, while the main panels show rescaled results in comparison to the theoretical expressions for the leading order term.
Symbols represent the most likely elongation of FENE-springs in MC simulations in the constant-{\em force} ensemble (see Appendix, Sec.~\ref{sec:zstar Methods}). 
}
\end{figure*}

\begin{figure*}
\includegraphics[width=\textwidth]{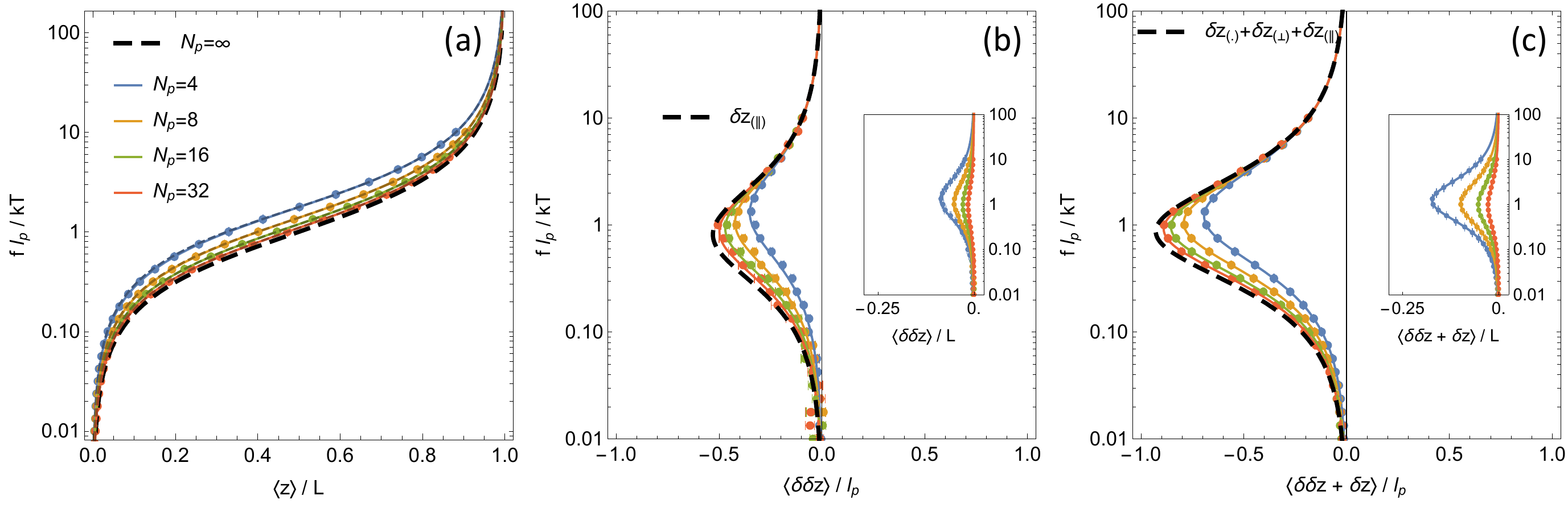}
\caption{
\label{fig:FENE z(f)}
FENE-springs in the constant-force ensemble:
(a)
elongation-force relations, 
(b)
finite-size corrections ($\langle \delta\delta z \rangle \equiv \langle z(f, \kappa) \rangle - \langle f(z, \kappa) \rangle^{-1}$) to the inverted force-elongation relations for chains of the same length, 
(c)
finite-size corrections ($\langle \delta\delta z + \delta z \rangle = \langle z(f, \kappa) \rangle - \langle z(f, \kappa=0) \rangle$) to the asymptotic elongation-force relation. 
Dashed colored lines in Panel (a) indicate the result of the Olver expansion, Eq.~(\ref{eq:QFENE Olver zeta(phi)}), while solid lines indicate the exact elongation-force relation, Eq.~(\ref{eq:QFENE z(f)}).
Symbols represent the average elongation of FENE-springs in MC simulations in the constant-force ensemble (see Appendix, Sec.~\ref{sec:spring simulation Methods}). 
Note that all results are shown with the dependent variable on the abscissa to simplify the comparison with Fig.~\ref{fig:FENE f(z)}.
}
\end{figure*}

\section{Stretching FENE-springs \label{sec:FENE}}
While it is reassuring to recover the well-known behavior of Gaussian chains, the model is too simple to provide a serious test of our approach.
In the following we explore the behavior of finitely extensible non-linear elastic (FENE)-springs~\cite{WarnerFENE1972} springs.
The radial distribution function for FENE springs, 
\begin{eqnarray}\label{eq:QFENE}
Q\left(\frac rL,N_p\right)
&\propto& \exp\left( \frac34 N_p \log\left[ 1-\left(\frac{r}{L}\right)^2\right] \strut\right)\nonumber\\
&\propto&\left( 1-\left(\frac{r}{L}\right)^2\right)^{\frac34 N_p }
\end{eqnarray}
reduces to the Gaussian distribution, Eq.~(\ref{eq:QG}), for $L\gg l_p$ as long as $r\ll L$. But contrary to Gaussian chains, the partition function of FENE-springs drops to zero in the limit of full elongation, $r\rightarrow L$.
As consequence, the contour length is a relevant independent length scale. 

The FENE model was not derived from an underlying microscopic chain model, but chosen for the relative ease with which it can be manipulated mathematically~\cite{WarnerFENE1972}.
Conveniently, the model can also be solved exactly in the present context (Sec.~\ref{sec:FENE exact}).
This provides us with a non-trivial test case for validating 
the ability of the approximation scheme outlined in the Theory Secs.~\ref{sec:ForceElongation} and~\ref{sec:ElongationForce} 
to predict the asymptotic behavior (Sec.~\ref{sec:FENE asymptotic}) as well as the leading order finite chain length corrections to the force-elongation and elongation-force relations (Secs.~\ref{sec:FENEfiniteForceEl} and \ref{sec:FENEfiniteElForce} respectively).
In addition, we use the FENE model as a stringent test case for validating the data analysis pipeline described in the Appendix 
(compare symbols to lines in Figs.~\ref{fig:FENE f(z)} and~\ref{fig:FENE z(f)}).

\subsection{Exact solution \label{sec:FENE exact}}
For FENE-springs, Eq.~(\ref{eq:QFENE}), all quantities of interest can be calculated exactly.
Integrating out transverse fluctuations yields
\begin{eqnarray}\label{eq:Z(z)FENE}
{\mathcal Z}(z/L,N_p)
&\propto&\left( 1-\left(\frac{z}{L}\right)^2\right)^{\frac34 N_p +1}\ .
\end{eqnarray}
Figure~\ref{fig:FENE f(z)}(a) illustrates the resulting force-elongation relation,
\begin{eqnarray}\label{eq:QFENE f(z)}
\frac{\langle f(z) \rangle}{k_BT/l_p}  =  \left(\frac32 + \frac{2}{N_p} \right) \frac{\left( \frac{z}{L} \right) } { 1 -  \left(\frac{z}{L}\right)^2 } \ ,
\end{eqnarray}
for a number of chain lengths, $N_p = 1/\kappa = 4,8,16,32$. As expected, the elastic response reduces to the Gaussian behavior for small elongations and diverges on approaching the limit of maximal elongation. There are discernable finite-size effects as shorter chains require a larger force to be constrained at a given relative elongation.

The partition function for the constant-force ensemble, Eq.~(\ref{eq:Partition function under force}), can also be calculated exactly and is given by
\begin{eqnarray}\label{eq:Z(f)FENE}
{\mathcal Z}\left(\frac{f l_p}{ k_BT}, N_p  \right)
&\propto& \frac{{\mathrm I}_{\frac32+\frac34 N_p} \left( N_p \frac{f l_p}{ k_BT} \right)}{\left( N_p \frac{f l_p}{ k_BT} \right)^{\frac32+\frac34 N_p}}
\end{eqnarray}
where ${\mathrm I}_\nu(x)$ denotes the modified Bessel function of the first kind and order $\nu$. 
By using Eqs.~(\ref{eq:DefineGf}) and~(\ref{eq:z(f) from G}) and employing the identity $x \, {\mathrm I}'_\nu(x) = x \, {\mathrm I}_{\nu+1}(x) + \nu \, {\mathrm I}_\nu(x)$, the corresponding elongation-force relation takes the form
\begin{eqnarray}\label{eq:QFENE z(f)}
\frac{\langle z(f) \rangle}{L}
&=& \frac {{\mathrm I}_{\frac52+\frac34 N_p}\left(N_p \frac{f l_p}{ k_BT}\ \right)} {{\mathrm I}_{\frac32+\frac34 N_p}\left(N_p \frac{f l_p}{ k_BT}\right)} \, .
\end{eqnarray}
Equations~(\ref{eq:Z(f)FENE}) and (\ref{eq:QFENE z(f)}) are difficult to interpret, since both, the order and the argument of the involved Bessel functions, depend on $N_p$.
Olver's uniform asymptotic expansion~\cite{Olver},
\begin{eqnarray}\label{eq:OlverExpansion}
I_\nu(x) & \sim & \frac{e^{\nu\eta(z)}}{(1+z^2)^{1/4}} \, , \nonumber\\
\eta(z) & = & (1+z^2)^{1/2} +\log \frac{z}{1+(1+z^2)^{1/2}} \, ,
\end{eqnarray}
depends with with $0< z \equiv x / \nu < +\infty$ on their {\em ratio} and helps to reduce the chain length dependence to a correction.
Substituting in Eq.~(\ref{eq:Z(f)FENE}), using dimensionless variables and differentiating yields
\begin{eqnarray}\label{eq:QFENE Olver zeta(phi)}
\langle \zeta(\phi) \rangle
&=&
\frac{\sqrt{9+36\kappa(1+\kappa)+16\phi^2}}{4\phi}\\
&&-\frac{3}{4\phi} \frac {9(1+2\kappa)^3+16(1+8\kappa/3)\phi^2}{9+36\kappa(1+\kappa)+16\phi^2} \, . \nonumber
\end{eqnarray}
Fig.~\ref{fig:FENE z(f)} shows elongation-force relations for the same chain lengths as in Fig.~\ref{fig:FENE f(z)}. The results are shown with the dependent variable on the abscissa to simplify the comparison with the force-elongation curves. The two sets of curves are qualitatively similar, but the finite-size effects are stronger for elongation-force relations.
For chains with $N_p\ge8$ the Olver approximation, Eq.~(\ref{eq:QFENE Olver zeta(phi)}), becomes virtually indistinguishable from the exact result, Eq.~(\ref{eq:QFENE z(f)}).

\subsection{Asymptotic behavior  \label{sec:FENE asymptotic}}
The asymptotic force-elongation relation for FENE-springs,
\begin{eqnarray}\label{eq:QFENE f0(z)}
\frac{ f(z/L)}{k_BT/l_p}  =  \frac32 \frac{\left( \frac{z}{L} \right) } {1- \left(\frac{z}{L}\right)^2} \ ,
\end{eqnarray}
can be read off straightforwardly from Eq.~(\ref{eq:QFENE f(z)}) and is indicated in Figs.~\ref{fig:FENE f(z)}(a) and \ref{fig:FENE z(f)}(a) as a dashed black line. In particular, with
\begin{equation}\label{eq:z(f) FENE large f}
\lim_{z\rightarrow L} \frac{f(z/L)}{k_BT/l_p} = \frac34  \frac1{1-z/L} 
\end{equation}
the elastic response diverges on approaching full elongation.

The same result, $f(z/L) = f_{(\cdot)}(z/L)$, also follows directly from $Q(r)$ via Eq.~(\ref{eq:fdot of Qp}) by neglecting fluctuations in the asymptotic limit
\begin{eqnarray}
\mathcal{Z}_{p,(\cdot)}(z/L) &=& \left( 1-\left(\frac{z}{L}\right)^2\right)^{\frac34}\ .
\end{eqnarray}
The inverse of the asymptotic force-elongation relation,
\begin{eqnarray}\label{eq:QFENE asymptotic z(f)}
\frac zL &=&
 \frac {\sqrt{9+ 16 \left( \frac{f l_p}{k_BT} \right)^2} } {4 \frac{ f l_p}{k_BT} }
-  \frac {3}4  \frac{k_BT}{ f l_p}\ .
\end{eqnarray}
agrees with Eq.~(\ref{eq:QFENE Olver zeta(phi)}) in the $\kappa\rightarrow0$-limit, where the Olver expansion becomes exact.

\subsection{Finite chain length corrections to the force-elongation relation}\label{sec:FENEfiniteForceEl}
The finite-size corrections to the force-elongation relation,
\begin{eqnarray}\label{eq:QFENE delta f(z)}
\frac{ \deltaf(z/L,N_p)}{k_BT/l_p}  =  \frac2{N_p} \frac{\left( \frac{z}{L} \right) } {1- \left(\frac{z}{L}\right)^2} \ ,
\end{eqnarray}
can again be read off straightforwardly from Eq.~(\ref{eq:QFENE f(z)}).
They turn out to be proportional to the asymptotic response and are shown for different chain lengths, $N_p$, in the inset of Fig.~\ref{fig:FENE f(z)}(b).
In particular, the corrections are linear in $\kappa$ with all higher order terms vanishing identically.
As a consequence, they perfectly superimpose, when they are rescaled as, $N_p \frac{ \deltaf(z/L)}{k_BT/l_p} = \frac{ \deltaf(z/L)}{k_BT/L}$ (Fig.~\ref{fig:FENE f(z)}(b)).

Following the analysis in Sec.~\ref{sec:ForceElongation Approximations} we can try to better understand the origin of the finite size corrections.
As in the case of Gaussian chains,  $\mathcal{Z}_{p,(\cdot)}(z/L,\kappa)  = \mathcal{Z}_{p,(\cdot)}(z/L)$ independently of chain length.
In the absence of finite size corrections to  the dominant free energy contribution, $\mathcal{F}_{p,(\cdot)}(z/L)$, there are also no corresponding corrections, Eq.~(\ref{eq:delta fdot of Qp}),  to the elastic response
\begin{eqnarray}
\frac{ \delta\!f_{(\cdot)}(z/L,N_p)}{k_BT/l_p} &=& 0\ .
\end{eqnarray}
However, for FENE-springs the effective spring constant, $k_{(\perp)}$, for transverse fluctuations diverges on approaching full elongation. The corresponding finite-size correction, Eq.~(\ref{eq:delta phi perp}), for the force-elongation curve reads
\begin{eqnarray}\label{eq:QFENE delta f perp(z)}
\frac{ \delta\!f_{(\perp)}(z/L,N_p)}{k_BT/l_p} &=&  \frac{2}{N_p} \frac{\left( \frac{z}{L} \right) } {1- \left(\frac{z}{L}\right)^2 } \ ,
\end{eqnarray}
so that indeed $f_{(\cdot)} + \delta\!f_{(\cdot)} + \delta\!f_{(\perp)} = \langle f \rangle$ for all values of $N_p$.

\subsection{Finite chain length corrections to the elongation-force relation}\label{sec:FENEfiniteElForce}
To first order in $\kappa$, the finite chain length corrections to the elongation-force relation Eq.~(\ref{eq:QFENE Olver zeta(phi)}) reads
\begin{equation}
\label{eq:FENE delta z}
\frac{\delta z(\phi,N_p)}{L} =
-\frac1{N_p}\left(
    \frac3{2 \phi} + \frac{8\phi}{9+16\phi^2} - \frac9{2\phi\sqrt{9+16\phi^2}}
 \right) \ .  
\end{equation}
Again, we can try to understand the origin of these finite size corrections following the analysis in Secs.~\ref{sec:ForceElongation Approximations} and~\ref{sec:ElongationForce}.
In Figs.~\ref{fig:FENE f(z)} and \ref{fig:FENE z(f)} we distinguish
(i) the difference between the inverted force-elongation relations for chains of finite length and the asymptotic elongation force relation (Fig.~\ref{fig:FENE f(z)}(c)),
(ii) the difference between the elongation-force relation and the inverted force-elongation relation for chains of a given length (Fig.~\ref{fig:FENE z(f)}(b)), and
(iii) the difference between the elongation-force relations for chains of finite length and  the asymptotic elongation-force relation (Fig.~\ref{fig:FENE z(f)}(c)), which are the sum of the first two terms.
In all three cases, insets show the absolute corrections, which are largest for short chains, while the main panels show rescaled corrections, $N_p (\delta z/L) = \delta z/l_p$.
All three corrections display qualitatively similar features. They are largest for chains, which are extended to about half of their maximal elongation, $z^\ast/L \approx 1/2$, and they vanish in the limits of small forces, $z^\ast/L \rightarrow 0$,  and of maximal elongation, $z^\ast/L \rightarrow 1$.

For a quantitative analysis,
consider first the finite-size corrections to the inverted force-elongation relation (Fig.~\ref{fig:FENE f(z)}(c)).
As expected, they converge to the sum of  the  first-order corrections arising from the dominant term and from transverse fluctuations, Eqs.~(\ref{eq:delta zeta dot}) and (\ref{eq:delta zeta perp}), which for FENE chains are given by
\begin{eqnarray}
\label{eq:FENE delta z dot}
\frac{\delta z_{(\cdot)}(z^\ast/L,N_p)}{L} &=&  0\\
\label{eq:FENE delta z perp}
\frac{\delta z_{(\perp)}(z^\ast/L,N_p)}{L} &=&  - \frac43 \, \frac1{N_p}\, \frac {z^\ast}L\, \ddfrac  { 1-  \left(\frac {z^\ast}L\right)^2} {1+  \left(\frac {z^\ast}L\right)^2}
\end{eqnarray}
and which we have indicated as a dashed black line in Fig.~\ref{fig:FENE f(z)}(c).
Note that there are higher order corrections to $\delta z$ even though $\deltaf$ is linear in $\kappa$, since the asymptotic force-elongation relation is non-linear.

Next consider the difference between the elongation-force relation, $\langle z(f)\rangle$ and the inverse of the force-elongation relation, $\langle f(z) \rangle$ (Fig.~\ref{fig:FENE z(f)}(b)).
In agreement with our theoretical arguments for the effect of elongation-dependent  longitudinal fluctuations, they converge to Eq.~(\ref{eq:delta zeta par}), which reads for FENE chains
\begin{equation}\label{eq:FENE delta z par}
\frac{\delta z_{(||)}(z^\ast/L,N_p)}{L} =  - \frac23\, \frac1{N_p}\, \frac {z^\ast}L\, \left( \left(\frac { 2} {  1+  \left(\frac {z^\ast}L\right)^2 }\right)^2 -1 \right)\ .
\end{equation}

Last, but not least, the total finite-size correction to the asymptotic elongation-force relation converges to the sum, $\delta z_{(\cdot)} + \delta z_{(\perp)}+ \delta z_{(||)}$, of the three correction terms (Fig.~\ref{fig:FENE z(f)}(c)). In particular, this sum can be shown to be equal Eq.~(\ref{eq:FENE delta z}) by using the asymptotic force-elongation relation, Eq.~(\ref{eq:QFENE f0(z)}), to express the stretching force through $z^\ast/L$.

\section{Stretching ``BTB-springs'' representing long WLC\label{sec:BTB}}

\begin{figure*}
\includegraphics[width=\textwidth]{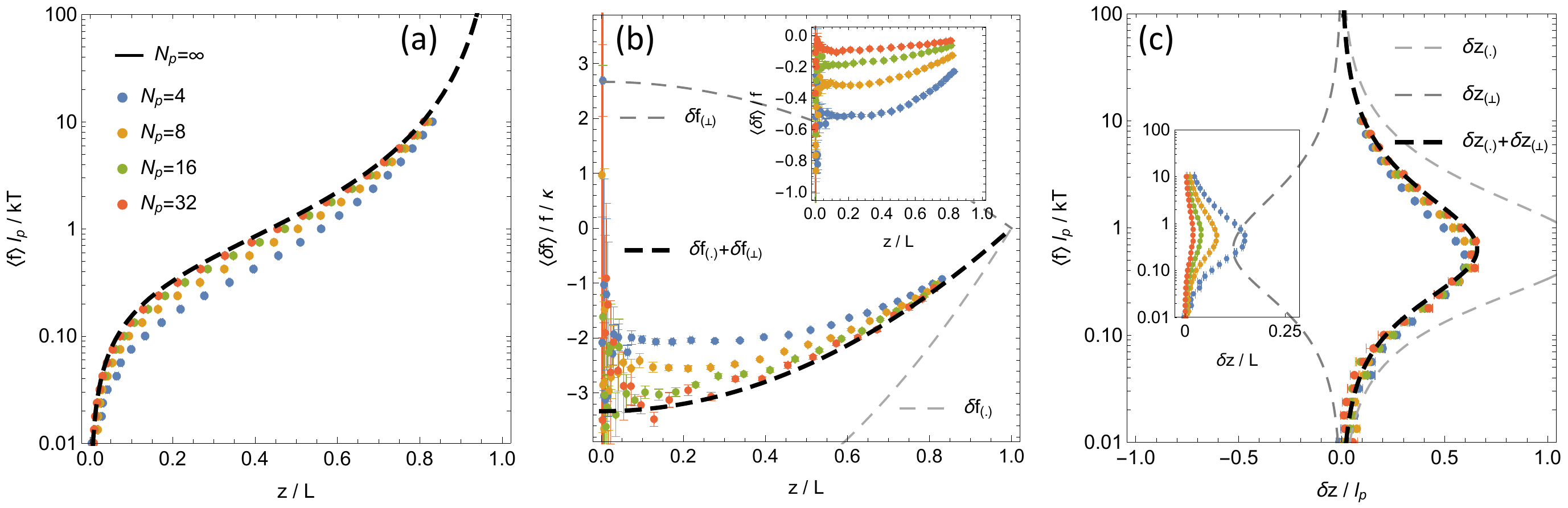}
\caption{
\label{fig:BTB f(z)}
BTB-springs in the constant-elongation ensemble:
(a)
force-elongation relations, 
(b)
finite-size corrections to the force-elongation relation, 
(c)
finite-size corrections to the inverted force-elongation relation. 
Symbols represent the most likely elongation of BTB-springs in MC simulations in the constant-{\em force} ensemble (see Appendix, Sec.~\ref{sec:zstar Methods}). 
Labels notation is as in Fig.~\ref{fig:FENE f(z)}.
}
\end{figure*}

\begin{figure*}
\includegraphics[width=\textwidth]{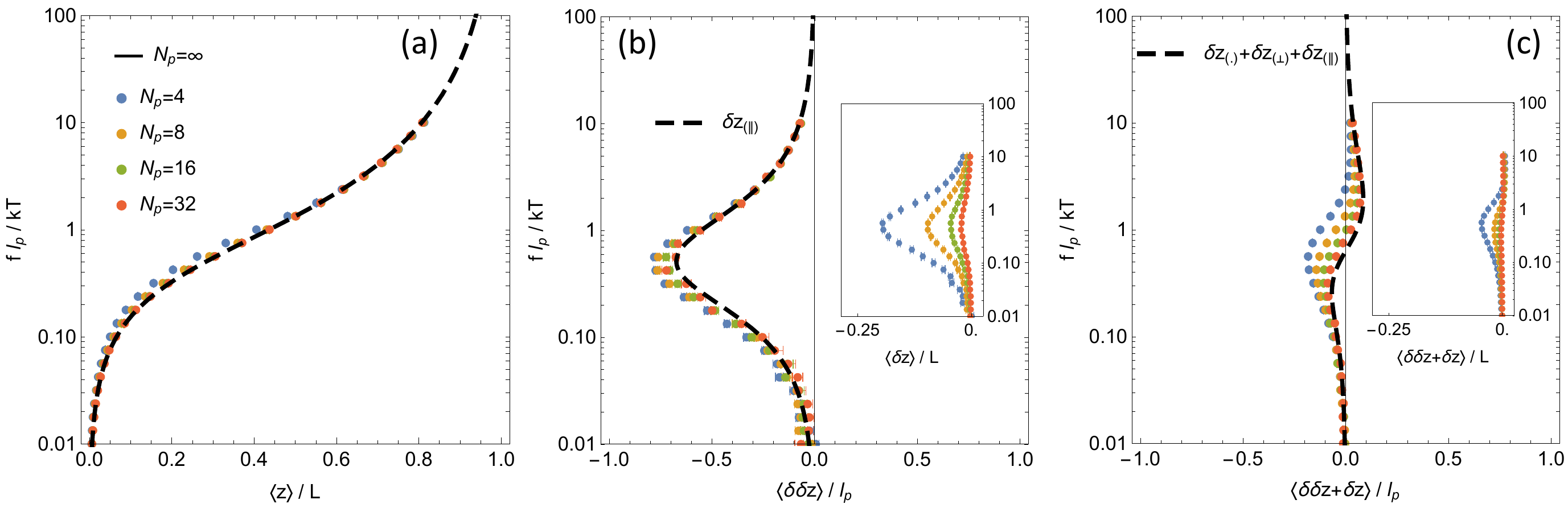}
\caption{
\label{fig:BTB z(f)}
BTB-springs in the constant-force ensemble:
(a)
elongation-force relations, 
(b)
finite-size corrections to the inverted force-elongation relations for chains of the same length, 
(c)
finite-size corrections to the asymptotic elongation-force relation. 
Symbols represent the average elongation of BTB-springs in MC simulations in the constant-force ensemble (see Appendix, Sec.~\ref{sec:spring simulation Methods}). 
Note that all results are shown with the dependent variable on the abscissa to simplify the comparison with Fig.~\ref{fig:BTB f(z)}.
Labels notation is as in Fig.~\ref{fig:FENE z(f)}.
}
\end{figure*}

In Ref.~\cite{BTBpdf-1997} Bhattacharjee, Thirumalai and Bryngelson used a variational approach~\cite{Ha1995,Ha1997} to replace the hard incompressibility constraint $\left|\frac{\partial}{\partial s}\vec r(s)\right|=1, \forall s\in[0,L]$ of the WLC by its thermal average $\langle (\frac{\partial}{\partial s}\vec r(s))^2 \rangle = 1$
and derived the following approximate formula for the end-to-end distribution function of a WLC chain:
\begin{equation}\label{eq:QBTB}
Q\left( \frac rL, N_p\right) \propto \left( 1 - \left(\frac rL \right)^2 \right)^{-9/2} \exp\left( -\frac34 N_p\frac1{1-(r/L)^2} \right) \, .
\end{equation}
Below we explore the properties of the corresponding BTB-springs.
As for FENE-springs, we can calculate their asymptotic behavior to first order in $\kappa=1/N_p$ using the approximation scheme outlined in the Theory Section~\ref{sec:Theory}. However, without an exact analytic solution, we are now limited to validating results for finite chain lengths through a comparison to numerical data from Monte Carlo simulations of stretched BTB-springs.
Figs.~\ref{fig:BTB f(z)} and~\ref{fig:BTB z(f)} for BTB-springs are the exact analogues of the Figs.~\ref{fig:FENE f(z)} and \ref{fig:FENE z(f)} for FENE-springs, which we have discussed in the preceding section.
For a comparison of BTB-springs to WLC see Sec.~\ref{sec:WLC} and the corresponding Figs.~\ref{fig:WLC f(z)} and~\ref{fig:WLC z(f)}.

\subsection{Asymptotic behavior \label{sec:BTB asymptotic}}
In the asymptotic limit, the free energy per persistence length is dominated by the exponential term in Eq.~(\ref{eq:QBTB}):
\begin{eqnarray}
\frac{\mathcal{F}_{p,(\cdot)}(z/L) }{k_BT}&=&
\frac34 \frac1{1-(r/L)^2}   \ .
\nonumber
\end{eqnarray}
Differentiating
with respect to the elongation, Eq.~(\ref{eq:fdot of Qp}), yields
\begin{equation}
\label{eq:QBTB fdot(z)}
\frac{ f_{(\cdot)} (z/L) }{k_BT/l_p} = \frac32 \frac{z/L } {\left(1- \left(z/L\right)^2\right)^2}
\end{equation}
for the asymptotic force-elongation relation of BTB-springs (shown as a dashed black line in Figs.~\ref{fig:BTB f(z)}(a) and~\ref{fig:BTB z(f)}(a)). 
BTB-springs display the same (Gaussian) small elongation behavior as FENE-springs, but their elastic response diverges more quickly on approaching full elongation:
\begin{equation}\label{eq:z(f) BTB large f}
\lim_{z\rightarrow L} \frac{f_{(\cdot)} (z/L)}{k_BT/l_p} = \frac38  \frac1{(1-z/L)^2} 
\end{equation}
Note that the limiting behavior perfectly agrees with the result of the corresponding direct variational calculation for stretched WLC~\cite{Ha1997}.
The closed expression Eq.~(\ref{eq:QBTB fdot(z)}) appears to be a new result.

The asymptotic elongation-force curve, $z_{(\cdot)}(f)$, can be expressed in closed form as a root of a third order polynomial. We nevertheless show results in the constant-force ensemble as parametric plots of the type $f(z^\ast)$ vs. $z(f(z^\ast))$ discussed in Sec.~\ref{sec:ElongationForce}.

\subsection{Finite chain length corrections to the force-elongation relation}\label{sec:BTBfiniteForceEl}

In contrast to FENE-springs, the relative finite chain length corrections to the force-elongation relation of BTP-springs are elongation dependent (Panels (a) and (b) in Figs.~\ref{fig:FENE f(z)} and \ref{fig:BTB f(z)}).
While they vanish close to full elongation, they are more than twice as strong for moderate elongations. In particular, they are of opposite sign.

For BTB-springs, the finite-size correction, Eq.~(\ref{eq:delta phi perp}), due to transverse fluctuations,
\begin{equation}\label{eq:QBTB delta f perp(z)}
\frac{\deltaf_{(\perp)}(z/L,\kappa) }{k_BT/l_p} =
4 \kappa \frac{z/L}{1-(z/L)^2} \ .
\end{equation}
has the exact same functional form as Eq.~(\ref{eq:QFENE delta f perp(z)}) for FENE springs, but is twice as strong.
In particular, Eqs.~(\ref{eq:QBTB delta f perp(z)}) and~(\ref{eq:QFENE delta f perp(z)}) have the same sign, since they result from the suppression of transverse fluctuations with increasing elongation.

The difference in behavior is due to the presence of finite-size corrections to the dominant free energy contribution, $\mathcal{F}_{p,(\cdot)}(z/L,\kappa)$, from aligned chains with the minimal elongation, $r=z$.
As there is no explicit chain length dependence in the subdominant prefactor in Eq.~(\ref{eq:QBTB}), there are no higher order corrections to $\delta\mathcal{F}_{p,(\cdot)}(z/L,\kappa)$ beyond the linear term,
\begin{eqnarray}\label{eq:QBTB delta Fp dot}
\frac{\delta\mathcal{F}_{p,(\cdot)}(z/L,\kappa) }{k_BT} =
  \frac92 \kappa \log\left(1-\left(z/L\right)^2 \right) \nonumber
\end{eqnarray}
Again the corresponding correction, Eq.~(\ref{eq:delta fdot of Qp}), for BTB-springs,
\begin{equation}\label{eq:QBTB delta f dot(z)}
\frac{\deltaf_{(\cdot)}(z/L,\kappa) }{k_BT/l_p} =
-9 \kappa \frac{z/L}{1-(z/L)^2} \ .
\end{equation}
has a FENE-like functional form. However, the sign is opposite, because the diverging subdominant factor of $\left(1-\left(\frac rL\right)^2 \right)^{-9/2}$ in Eq.~(\ref{eq:QBTB}) {\em reduces} the drop in $Q(r)$ on approaching full elongation. A comparison of the prefactors with Eq.~(\ref{eq:QBTB delta f perp(z)}) shows, that this latter effect is larger and hence the overall correction of opposite sign compared to FENE-springs.
Our numerical results for BTB-springs are in excellent agreement with this analysis (Fig.~\ref{fig:BTB f(z)}(b)).

\subsection{Finite chain length corrections to the elongation-force relation}\label{sec:BTBfiniteElForce}

Following the discussions in Secs.~\ref{sec:ForceElongation} and~\ref{sec:ElongationForce}, 
the finite-chain length corrections to the asymptotic force-elongation relation also cause first-order corrections to its inverse. For BTB-springs Eqs.~(\ref{eq:delta zeta dot}) and (\ref{eq:delta zeta perp}) read:
\begin{eqnarray}
\label{eq:BTB delta z dot}
\frac{\delta z_{(\cdot)}(z^\ast/L,\kappa)}{L} &=&  +6 \kappa \frac{(z^\ast/L) \left( 1-(z^\ast/L)^2\right)^2} {1+3(z^\ast/L)^2}\\
\label{eq:BTB delta z perp}
\frac{\delta z_{(\perp)}(z^\ast/L,\kappa)}{L} &=& -\frac83 \kappa \frac{(z^\ast/L) \left( 1-(z^\ast/L)^2\right)^2} {1+3(z^\ast/L)^2} \ .
\end{eqnarray}
The two functions and their sum are shown in Fig.~\ref{fig:BTB f(z)}(c).
Like for FENE-springs, the corrections are strongest around $z/L \approx 1/2$ and $f l_p/k_BT \approx 1$. 
However, the total correction has the opposite sign and its magnitude  is about 50\% larger.
Once more, the numerical results for BTB-springs are in excellent agreement with our analysis.

In addition, we expect a correction due to the elongation-dependence of longitudinal fluctuations. For BTB-springs Eq.~(\ref{eq:delta zeta par}) reads
\begin{eqnarray}
\label{eq:BTB delta z par}
\lefteqn{\frac{\delta z_{(||)}(z^\ast/L,N_p)}{L} =}\\
&& -4 \kappa \frac{\left( (z^\ast/L) +  (z^\ast/L)^3 \right) \left( 1-(z^\ast/L)^2\right)^2} {\left(1+3(z^\ast/L)^2\right)^2} \ .\nonumber
\end{eqnarray}
The correction is qualitatively similar, but stronger than for FENE-springs (panels (b) in Figs.~\ref{fig:FENE z(f)} and \ref{fig:BTB z(f)}). Again our numerical results converge to the theoretical prediction.

The total finite chain length corrections to the elongation relation of  BTB-springs are shown in Fig.~\ref{fig:BTB z(f)}(c).
Curiously, the theoretically predicted first order correction, $\delta z_{(\cdot)}+\delta z_{(\perp)}+\delta z_{(||)}$, for BTB-springs almost cancel each other. 
Once more the results of our simulations for BTB-springs representing WLCs with a length $N_p=4, \ldots, 32$ persistence lengths converge to the theoretically predicted first order correction.

\section{Stretching ``BRE-springs'' representing long WLC\label{sec:BRE}}

\begin{figure*}
\includegraphics[width=\textwidth]{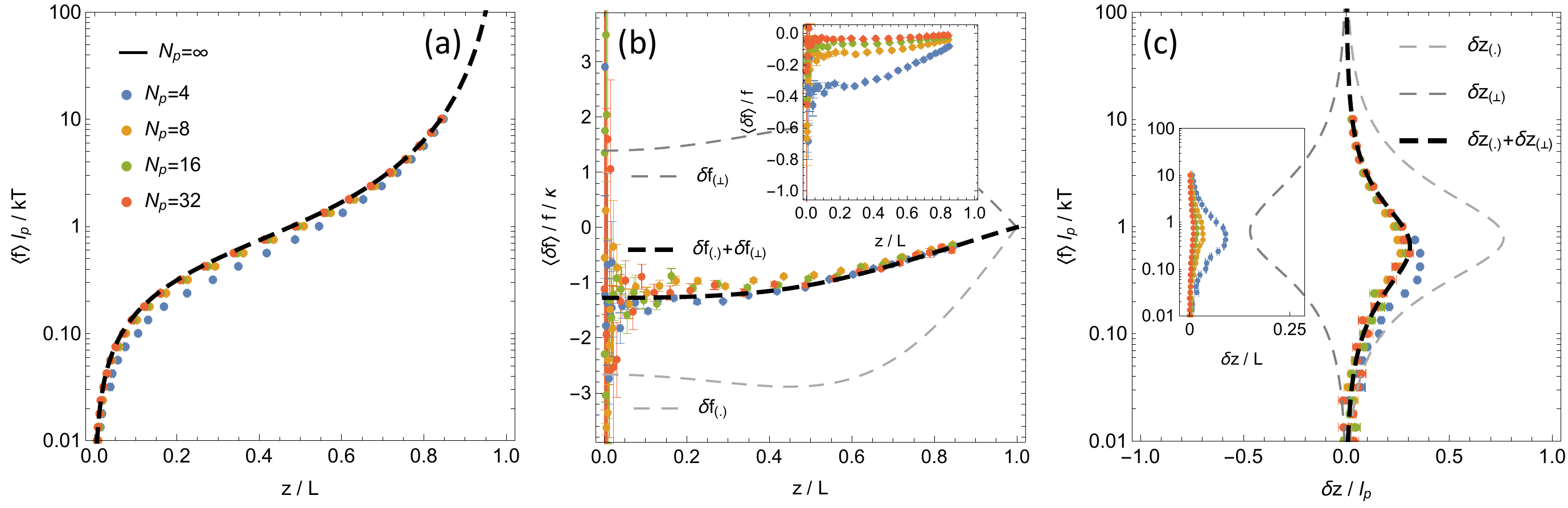}
\caption{
\label{fig:BRE f(z)}
BRE-springs in the constant-elongation ensemble:
(a)
force-elongation relations, 
(b)
finite-size corrections to the force-elongation relation, 
(c)
finite-size corrections to the inverted force-elongation relation. 
Symbols represent the most likely elongation of BRE-springs in MC simulations in the constant-{\em force} ensemble (see Appendix, Sec.~\ref{sec:zstar Methods}). 
Labels notation is as in Fig.~\ref{fig:FENE f(z)}.
}
\end{figure*}

\begin{figure*}
\includegraphics[width=\textwidth]{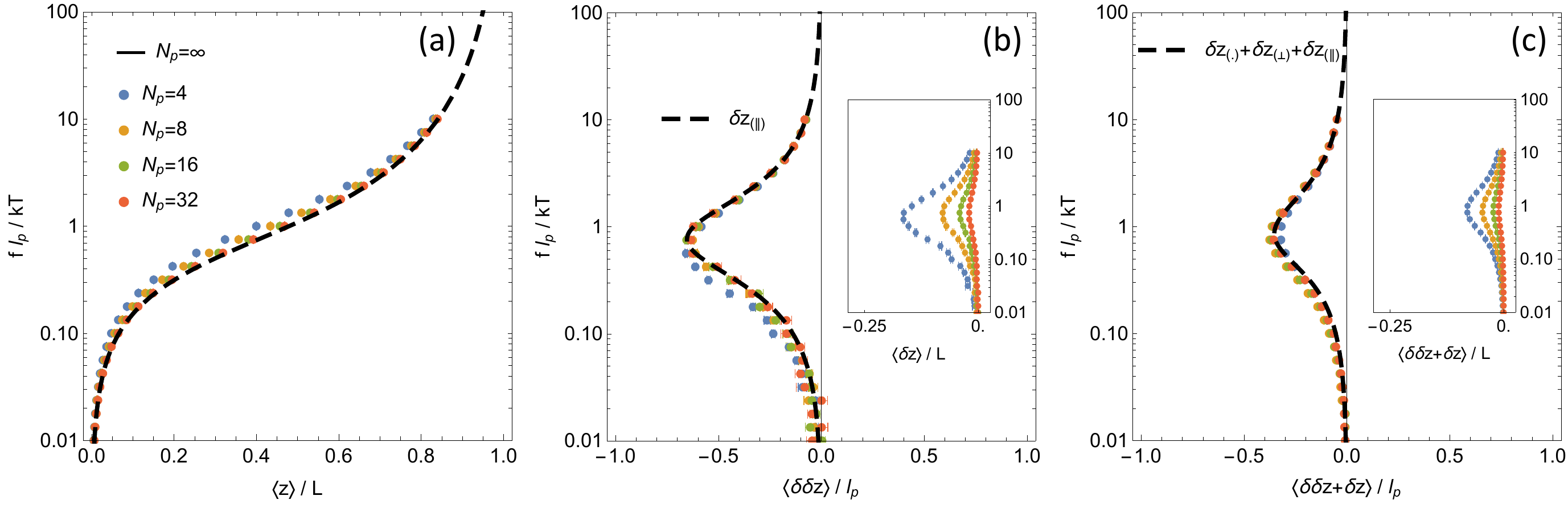}
\caption{
\label{fig:BRE z(f)}
BRE-springs in the constant-force ensemble:
(a)
elongation-force relations, 
(b)
finite-size corrections to the inverted force-elongation relations for chains of the same length, 
(c)
finite-size corrections to the asymptotic elongation-force relation. 
Symbols represent the average elongation of BRE-springs in MC simulations in the constant-force ensemble (see Appendix, Sec.~\ref{sec:spring simulation Methods}). 
Note that all results are shown with the dependent variable on the abscissa to simplify the comparison with Fig.~\ref{fig:BRE f(z)}.
Labels notation is as in Fig.~\ref{fig:FENE z(f)}.
}
\end{figure*}

In Ref.~\cite{BeckerRosaEveraers2010} we have carried out a systematic evaluation of the quality of available analytical expressions for  the end-to-end distance distribution of WLC.
In particular, we have proposed a closed analytical expression,
\begin{eqnarray}\label{eq:QBRE}
Q_{\rm BRE}(r) \equiv ( 1 - c r^2 )^{5/2} \, Q_A(r) \, Q_B(r) \, , 
\end{eqnarray}
for the end-to-end distance distribution of WLC composed of three factors, which interpolates between all relevant limiting cases from stiff to flexible chains and from looped to fully stretched configurations.
In analogy to the FENE-case, $\mathcal{F}_{\rm BRE}(r) = -k_BT \log\left(Q_{\rm BRE}(r) \right)$ describes the elastic (free) energy of a non-linear, finite-extensible spring.
For notational conciseness and want of a better name we will refer to corresponding results as describing the behavior of ``BRE-springs''. 

In the context of DNA stretching, we are mostly interested in chains, which are much longer than their persistence length, $L\ge 8 l_p$.
In this case, $c\approx 0$ and $Q_B(r)\propto1$ 
and we can extract all distance dependent factors of the radial distribution function from
\begin{eqnarray}\label{eq:QA}
Q_A(r)
&\propto& \left(1-\left(\frac rL\right)^2 \right)^{-5/2} \times \\
             && \exp\left( \frac{-\frac12 \left(\frac rL\right)^2 + \frac{17}{16}\left(\frac rL\right)^4 - \frac{9}{16}\left(\frac rL\right)^6}{1-\left(\frac rL\right)^2}       \right)\times\nonumber\\
             && \exp\left( \frac{-\frac34 \left(\frac rL\right)^2 + \frac{23}{64}\left(\frac rL\right)^4 - \frac{7}{64}\left(\frac rL\right)^6}{1-\left(\frac rL\right)^2}    \frac{L}{l_p}   \right)\nonumber
\end{eqnarray}
which we obtained~\cite{BeckerRosaEveraers2010}  by a systematic interpolation between the exact limit results by Daniels~\cite{Daniels1952} and Wilhelm and Frey~\cite{WilhelmFreyRadialWLCPRL1996}.

As in the case of BTB-springs, we first explore the properties of BRE-springs as such. Figs.~\ref{fig:BRE f(z)} and \ref{fig:BRE z(f)} for BRE-springs are the exact analogues of  Figs.~\ref{fig:FENE f(z)} and \ref{fig:FENE z(f)} for FENE-springs and Figs.~\ref{fig:BTB f(z)} and \ref{fig:BTB z(f)} for BTB-springs. Again we have not been able to obtain an exact analytic solution. As a consequence, we are restricted to validating results obtained from the approximation scheme outlined in the Theory Section~\ref{sec:Theory} via a comparison to numerical data from Monte Carlo simulations of stretched BRE-springs.
For a comparison of BRE-springs to WLC see Sec.~\ref{sec:WLC} and the corresponding Figs.~\ref{fig:WLC f(z)} and \ref{fig:WLC z(f)}.

\subsection{Asymptotic behavior \label{sec:BRE asymptotic}}
In the asymptotic limit, the free energy per persistence length is given by the dominant exponential term in Eq.~(\ref{eq:QA}):
\begin{eqnarray}
\frac{\mathcal{F}_{p,(\cdot)}(z/L) }{k_BT}&=&
-\frac{-\frac34 \left(z/L\right)^2 + \frac{23}{64}\left(z/L\right)^4 - \frac{7}{64}\left(z/L\right)^6}{1-\left(z/L\right)^2}  \ .
\nonumber
\end{eqnarray}
Differentiating
with respect to the elongation, Eq.~(\ref{eq:fdot of Qp}), yields
\begin{equation}
\label{eq:QA fdot(z)}
\frac{ f_{(\cdot)} (z/L) }{k_BT/l_p} = \frac12 \frac zL + \frac{z/L}{\left(1-(z/L)^2\right)^{2}} -\frac7{16} \left(\frac zL\right)^3
\end{equation}
for the asymptotic force-elongation relation of BRE-springs. 
The behavior is similar to BTB-springs and largely dominated by the first two terms, which reproduce the exactly know behavior of WLC in the two limits of weak and strong elongation, Eqs.~(\ref{eq:z(f) WLC large f}) and (\ref{eq:z(f) WLC small f}), respectively.
The inverse, $z_{(\cdot)}(f)$, being the root of a seventh order polynomial, we show results in the constant-force ensemble as parametric plots.

\begin{figure*}
\includegraphics[width=\textwidth]{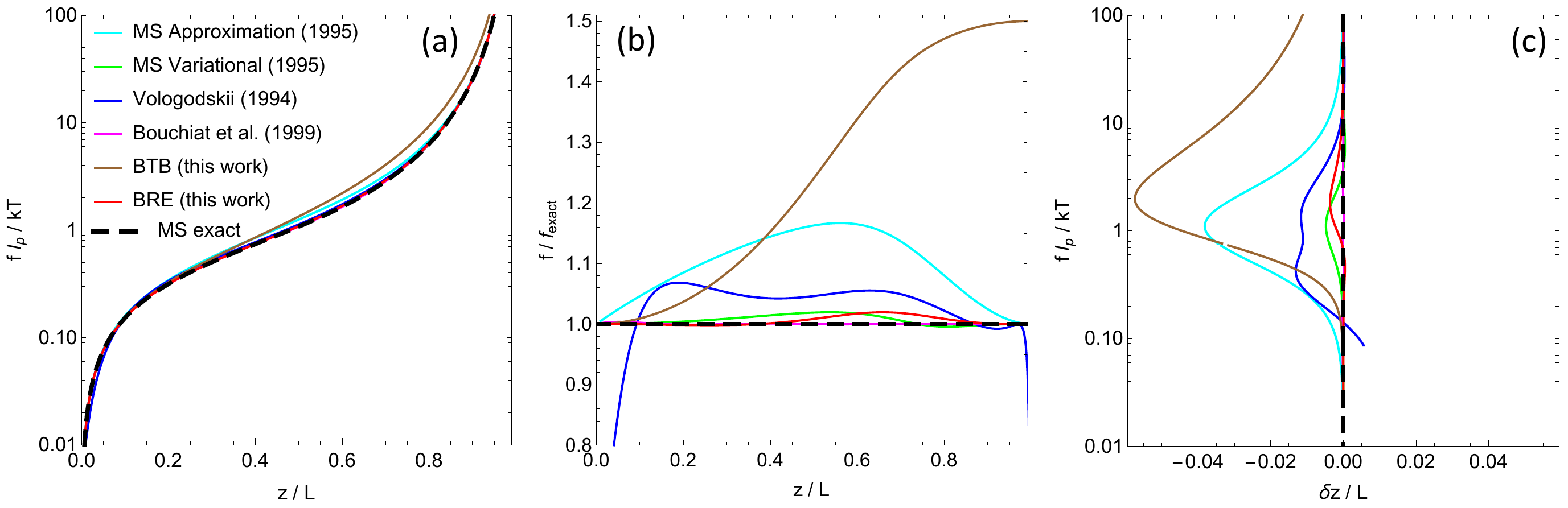}
\caption{\label{fig:WLC Asymptotics}
(a)
Asymptotic force-elongation relation for WLC, 
(b)
relative error of predicted forces, 
(c)
error of predicted position. 
Black line: Marko's and Siggia's exact solution.
Cyan line: Marko's and Siggia's approximate expression Eq.~(\ref{eq:MarkoSiggia f(z)}).
Green line: numerical solution of the Marko and Siggia variational theory.
Blue line: Vologodskii's approximate expression~\cite{Vologodskii1994}.
Magenta line: Bouchiat {\it et al.}'s approximate expression~\cite{BouchiatCroquetteBJ1999}.
Brown and red lines: analytical expressions, Eq.~(\ref{eq:QBTB fdot(z)}) and~(\ref{eq:QA fdot(z)}), derived from the BTB~\cite{BTBpdf-1997} and BRE~\cite{BeckerRosaEveraers2010} distributions.
}
\end{figure*}

\begin{figure*}[h]
\includegraphics[width=\textwidth]{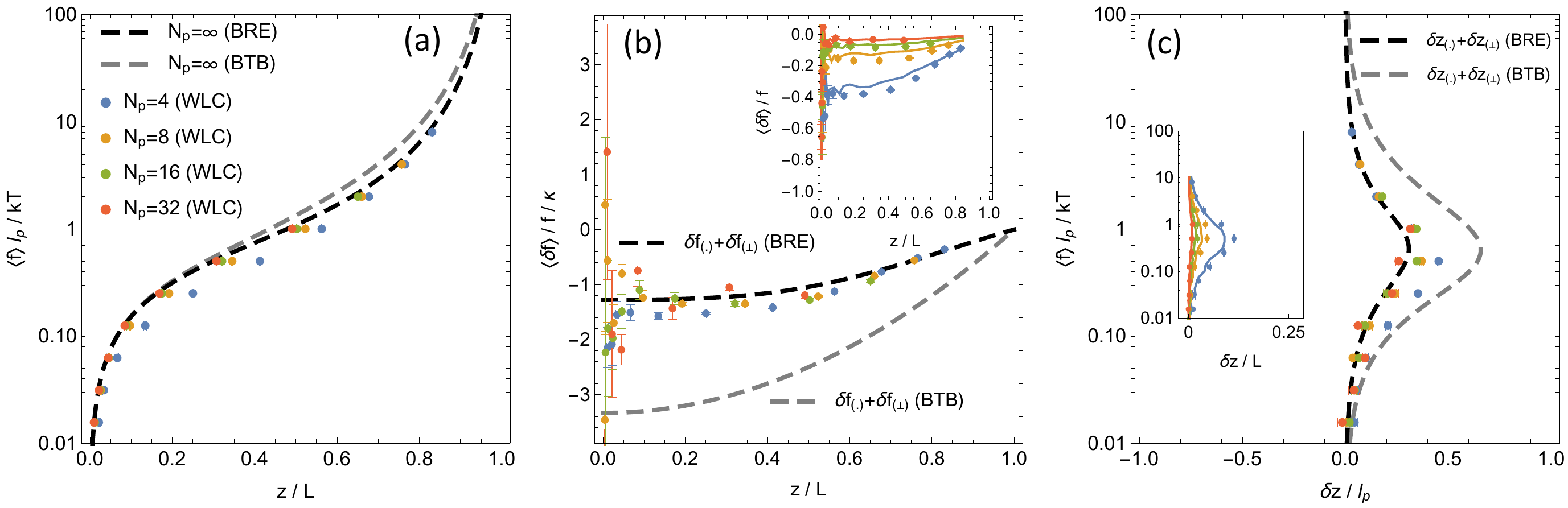}
\caption{
\label{fig:WLC f(z)}
WLC of finite length compared to corresponding BTB- and BRE-springs in the constant-elongation ensemble:
(a)
force-elongation relations, 
(b)
finite-size corrections to the force-elongation relation, 
(c)
finite-size corrections to the inverted force-elongation relation. 
Symbols:  most likely elongations of WLC in MC simulations in the constant-{\em force} ensemble (see Appendix, Sec.~\ref{sec:zstar Methods}). 
Gray dashed lines: theoretical results for BTB-springs (see Fig.~\ref{fig:BTB f(z)}).
Black dashed lines: theoretical results for BRE-springs (see Fig.~\ref{fig:BRE f(z)}).
Solid colored lines in the insets: numerical results for BRE-springs representing WLC of finite length (see Fig.~\ref{fig:BRE f(z)}).
Colors distinguish chain lengths,
label notation is as in Fig.~\ref{fig:FENE f(z)}.
}
\end{figure*}

\begin{figure*}[h]
\includegraphics[width=\textwidth]{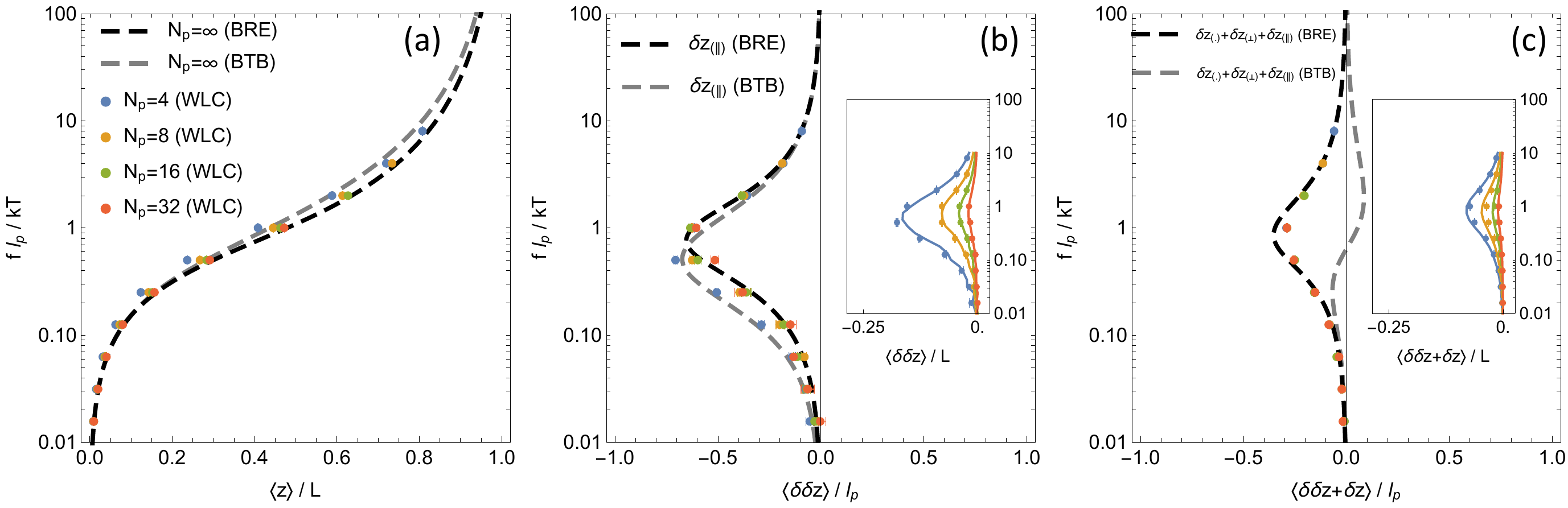}
\caption{
\label{fig:WLC z(f)}
WLC of finite length compared to corresponding BTB- and BRE-springs in the constant-force ensemble:
(a)
elongation-force relations, 
(b)
finite-size corrections to the inverted force-elongation relations for chains of the same length, 
(c)
finite-size corrections to the asymptotic elongation-force relation. 
Symbols: average elongation of WLC in MC simulations in the constant-force ensemble (see Appendix, Sec.~\ref{sec:WLC simulation Methods}). 
Gray dashed lines: theoretical results for BTB-springs (see Fig.~\ref{fig:BTB z(f)}).
Black dashed lines: theoretical results for BRE-springs (see Fig.~\ref{fig:BRE z(f)}).
Solid colored lines in the insets: numerical results for BRE-springs representing WLC of finite length (see Fig.~\ref{fig:BRE z(f)}).
Colors distinguish chain lengths,
label notation is as in Fig.~\ref{fig:FENE z(f)}.
}
\end{figure*}

\subsection{Finite chain length corrections to the force-elongation relation}\label{sec:BREfiniteForceEl}
As in the other cases, the effective spring constant, Eq.~(\ref{eq:kperp}), for transverse fluctuations of BRE-springs,
\begin{eqnarray}
k_{(\perp)}(z/L)
\propto \frac12 + \frac{1}{\left(1-(z/L)^2\right)^{2}} -\frac7{16} \left(\frac zL\right)^2\ ,
 \end{eqnarray}
diverges on approaching full elongation. To first order in $\kappa$ the corresponding finite-size correction, Eq.~(\ref{eq:delta phi perp}), for the force-elongation curve reads
\begin{equation}\label{eq:QA delta f perp(z)}
\frac{\deltaf_{(\perp)}(z/L,\kappa) }{k_BT/l_p} =
\kappa\ \ddfrac
  { \frac zL \left(   \frac{4}{\left(1-(z/L)^2\right)^{3}} -\frac{7}{8} \right)}
  {\frac12 +  \frac{1}{\left(1-(z/L)^2\right)^{2}} -\frac{7}{16} \left(\frac zL\right)^2}\ .
\end{equation}
While the effect of transverse fluctuations is qualitatively similar in all three cases (Panel (b) in Figs.~\ref{fig:FENE f(z)}, \ref{fig:BTB f(z)} and \ref{fig:BRE f(z)}), the corrections for BRE-springs are somewhat smaller than for BTB-springs.

Similarly to BTB-springs, BRE-springs exhibit finite-size corrections to the dominant free energy contribution, $\mathcal{F}_{p,(\cdot)}(z/L,\kappa)$, from aligned chains with the minimal elongation, $r=z$.
Again there is no explicit chain length dependence in the two subdominant factors in Eq.~(\ref{eq:QA}). As a consequence, there are no higher order corrections to $\delta\mathcal{F}_{p,(\cdot)}(z/L,\kappa)$ beyond the linear term,
\begin{eqnarray}\label{eq:QA delta Fp dot}
\lefteqn{ \frac{\delta\mathcal{F}_{p,(\cdot)}(z/L,\kappa) }{k_BT} =
\kappa \left(
  \frac52 \log\left(1-\left(z/L\right)^2 \right) \right.}\\
&& \left.
  -\frac{-\frac12 \left(z/L\right)^2 + \frac{17}{16}\left(z/L\right)^4 - \frac{9}{16}\left(z/L\right)^6}{1-\left(z/L\right)^2}
\right) \ .
\nonumber
\end{eqnarray}
The corresponding correction, Eq.~(\ref{eq:delta fdot of Qp}), to the elastic response reads
\begin{equation}\label{eq:QA delta f dot(z)}
\frac{\deltaf_{(\cdot)}(z/L,\kappa) }{k_BT/l_p} =
\kappa \left(
\frac zL - 5\frac{z/L}{1-(z/L)^2} -\frac9{4} \left(\frac zL\right)^3
\right)\ .
\end{equation}
Note that the dominant FENE-like term in Eq.~(\ref{eq:QA delta f dot(z)}) has again the opposite sign from Eq.~(\ref{eq:QFENE delta f(z)}).
As for BTB-springs the diverging subdominant factor of $\left(1-\left(\frac rL\right)^2 \right)^{-5/2}$ in Eq.~(\ref{eq:QA}) {\em reduces} the drop in $Q_A(r)$ on approaching full elongation. Because of the smaller exponent, this correction is again smaller for BRE- than for BTB-springs.

For BRE-springs the sum of the two corrections is approximately given by the more readable expression
\begin{equation}
\label{eq:First order correction to f(z)}
\frac{ \deltaf(z/L,\kappa)}{k_BT/l_p}  \approx \kappa \left( \frac12 -\frac1{2\left(1-z/L\right)}-\frac{17}{12}\frac zL\right)\ .
\end{equation}
While the total correction is qualitatively similar to the one for BTB-springs, it turns out to be only about half as strong. Compared to FENE-springs, the major difference is again the opposite sign caused  by $\deltaf_{(\cdot)}(z/L,\kappa)$.
Our numerical results for BRE-springs are in excellent agreement with this analysis (Fig.~\ref{fig:BRE f(z)}(b)).

\subsection{Finite chain length corrections to the elongation-force relation}\label{sec:BREfiniteElForce}
In the constant-elongation ensemble, we expect a correction due to the elongation-dependence of longitudinal fluctuations, because the corresponding effective spring constant,
\begin{eqnarray}\label{eq:k_|| BRE}
k_{(||)}(z/L)
&\propto&   \frac{ f_{(\cdot)}' (z/L) }{k_BT/l_p} \\
&=& \frac12 + \frac{1+3(z/L)^2}{\left(1-(z/L)^2\right)^{3}} -\frac{21}{16} \left(\frac zL\right)^2\ ,
 \end{eqnarray}
diverges even more rapidly than $k_{(\perp)}(z/L) $.  For BRE-springs Eq.~(\ref{eq:delta zeta par}) reads
\begin{eqnarray}
\label{eq:WLC delta z par}
\lefteqn{\frac{\delta z_{(||)}(z^\ast/L,N_p)}{L} = 48 \kappa (z^\ast/L) \left(1-(z/L)^2\right)^{2} \times}\\
&&\ddfrac {-25 -60 (z^\ast/L)^2 + 42 (z^\ast/L)^4 - 28 (z^\ast/L)^6 + 7 (z^\ast/L)^8 }
            {\left(24 +3  (z^\ast/L)^2 + 87 (z^\ast/L)^4 - 71 (z^\ast/L)^6 + 21 (z^\ast/L)^8  \right)^2}\nonumber
\end{eqnarray}
and can be approximated as
\begin{eqnarray}
\lefteqn{\frac{\delta z_{(||)}(z^\ast/L,N_p)}{L} } \\
&\approx& \frac{l_p}{2L} / \sqrt{\left(\frac{6}{25}\frac L{z^\ast}  \right)^2 + \left(\frac16 \frac1{(1-z^\ast/L)^2}  \right)^2}      \nonumber
\end{eqnarray}
The behavior shown in Fig.~\ref{fig:BRE z(f)}(b) is very similar to the results for the other cases. Again our numerical results converge to the theoretical prediction.

Following the discussions in Sec.~\ref{sec:ElongationForce}, 
we need to add $\delta z_{(||)}$ to the finite-chain length corrections from the inverted force-elongation relation given by Eqs.~(\ref{eq:delta zeta dot}) and (\ref{eq:delta zeta perp}).
With $\deltaf_{(\perp)}$, $\deltaf_{(\cdot)}$ and $f_{(\cdot)}'$ for BRE-springs defined in Eqs.~(\ref{eq:QA delta f perp(z)}), (\ref{eq:QA delta f dot(z)}) and (\ref{eq:k_|| BRE}), the expressions
\begin{eqnarray}
\label{eq:WLC delta z perp}
\frac{\delta z_{(\perp)}(z^\ast/L,\kappa)}{L} &=&   -\frac{\deltaf_{(\perp)}(z^\ast/L,\kappa)}{f_{(\cdot)}'(z^\ast/L)}\\
\label{eq:WLC delta z dot}
\frac{\delta z_{(\cdot)}(z^\ast/L,\kappa)}{L} &=&  -\frac{\deltaf_{(\cdot)}(z^\ast/L,\kappa)}{f_{(\cdot)}'(z^\ast/L)}\
\end{eqnarray}
gain little in being written out in full.
The two functions are shown in Fig.~\ref{fig:BRE f(z)}(c).
Like in the other cases, the corrections are strongest around $z/L \approx 1/2$ and $f l_p/k_BT \approx 1$. 
As for BTB-springs, $\delta z_{(\cdot)}$ and $\delta z_{(\perp)}$ have opposite signs. But with $\delta z_{(\cdot)}$ being smaller, the sum, $\delta z_{(\cdot)}+\delta z_{(\perp)}$, is about 50\% smaller. Again, the numerical results for BRE-springs are in excellent agreement with our analysis.

The total finite chain length corrections to the elongation-force relation of BRE-springs are shown in Fig.~\ref{fig:BRE z(f)}(c).
The first point to note is again the excellent agreement between the results of our simulations for chains with a length $N_p=4, \ldots, 32$ persistence lengths and the theoretically predicted first order correction, $\delta z_{(\cdot)}+\delta z_{(\perp)}+\delta z_{(||)}$. Higher order terms appear to be negligible.
A second key feature is revealed in the direct comparison to the corresponding Figs.~\ref{fig:FENE z(f)}(c) and \ref{fig:BTB z(f)}(c) for FENE- and BTB-springs: due to the magnitude and opposite sign of the contribution $\delta z_{(\cdot)}$, the total finite chain length corrections to the elongation-force relation of BRE-springs are surprisingly small, even though they do not exhibit the near cancellation we found in the case of BTB-springs.

\section{Discussion: Stretching wormlike chains}\label{sec:WLC}

In the present paper, we have developed a formalism for inferring force-elongation and elongation-force relations for single-molecule stretching experiments from given (approximate) expressions for the chain end-to-end distance distribution.
We have validated the formalism for the analytically exactly solvable case of FENE-springs (Sec.~\ref{sec:FENE}).
In Secs.~\ref{sec:BTB}  and \ref{sec:BRE} we have derived the relevant expressions for the approximate BTB- and BRE-distributions for long WLC, whose contour length is much larger than their persistence length, $L/l_p=N_p\gg1$.

We now turn to the question which if any of the approximate radial distribution functions allows us to derive a quantitative description of the behavior of wormlike chains.
We will follow the same outline as in the preceding sections.
In Section~\ref{sec:WLC Asymptotic} we compare the asymptotic force-elongation relation of BTB- and BRE-springs to the results of Marko and Siggia.
In the second step (Sec.~\ref{sec:ComparisonToSimuls}), we use our numerical results for WLC  to test the corresponding expressions for the finite-chain length corrections.
As a final point, we show in Section~\ref{sec:Nicks} how experimentalists might employ our results to infer the changing number of nicks in a ds-DNA molecule by observing the changing mean elongation in a single-molecule stretching experiment, where the DNA is held at constant force.

\subsection{Asymptotic behavior}\label{sec:WLC Asymptotic}

In Fig.~\ref{fig:WLC Asymptotics} we compare the asymptotic force-elongation relation for BTB- and BRE-springs, Eqs.~(\ref{eq:QBTB fdot(z)}) and (\ref{eq:QA fdot(z)}), to
the MS approximate expression Eq.~(\ref{eq:MarkoSiggia f(z)}),
the numerical solution of the MS variational theory,
an analytical expressions proposed by Vologodskii (Eq.~(4) in Ref.~\cite{Vologodskii1994}),
the exact MS solution obtained by numerically inverting a $100\times100$ matrix, and
an empirical formula by Bouchiat {\it et al.} (Eq.~(11) in Ref.~\cite{BouchiatCroquetteBJ1999}), who fitted a seventh order polynomial to the difference between the exact solution and Eq.~(\ref{eq:MarkoSiggia f(z)}).
Considered over the full force and elongation range in Panel (a), all approximate expressions provide a good approximation to the exact solution (indicated by a dashed black line).
Only for Eq.~(\ref{eq:QBTB fdot(z)}) the deviations are immediately apparent.

For a more detailed analysis we have calculated the relative error of the asymptotic force-elongation relations (Panel (b)) and the absolute error of the asymptotic elongation-force relations (Panel (c)).
These representations show that the deviations of the BRE-spring expression, Eq.~(\ref{eq:QA fdot(z)}), are of the order of 2\% over the full range of elongations.
They are thus about one order of magnitude smaller than for the MS approximation, Eq.~(\ref{eq:MarkoSiggia f(z)}) and comparable to the numerical evaluation of the MS variational theory. Vologodskii's expression is three to five times worse in the intermediate force regime and breaks down in both limits~\cite{BouchiatCroquetteBJ1999}.
For BTB-springs the elastic response to large forces is off by a factor of 3/2~\cite{Ha1997}.
While Bouchiat {\it et al.}'s~\cite{BouchiatCroquetteBJ1999} fit of the exact MS solution retains its utility for the analysis of experimental data, Eq.~(\ref{eq:QA fdot(z)}) has at least the merit of being the most precise explicit expression resulting from a systematic theoretical approach to the problem.

\subsection{Finite chain length effects}\label{sec:ComparisonToSimuls}
In the absence of exact results for the finite chain length corrections to the force-elongation and elongation-force relation of WLC, we are limited to comparing the predictions we have derived from the BTB- and BRE-distributions to our numerical data for WLC.
In order not to confuse the errors in the inferred asymptotic force-elongation relations with the predicted finite-chain length corrections, we calculate the latter for our WLC data relative to the exact asymptotic MS force-elongation relation.

The presentation of our results in Figs.~\ref{fig:WLC f(z)} and \ref{fig:WLC z(f)} is the exact analogue of Figs.~\ref{fig:BTB f(z)} and \ref{fig:BTB z(f)} for BTB-springs and of Figs.~\ref{fig:BRE f(z)} and \ref{fig:BRE z(f)} for BRE-springs.
The only difference is that symbols now represent simulation results for WLC, while gray and black dashed lines represent predictions for BTB- and BRE-springs respectively. 
In the constant-force ensemble, there is excellent agreement between the WLC data and the finite chain length corrections inferred from the BRE-distribution.
In contrast, the BTB-results -- while qualitatively perfectly reasonable -- are off by a factor of 2-3 over the entire range of elongations (Fig.~\ref{fig:WLC f(z)}(b) and (c)). 
We tentatively conclude, that the predicted asymptotic force-elongation relations and finite chain lengths corrections seem to be of comparable quality.
In the constant-elongation ensemble, there is very good agreement between the observed corrections due to longitudinal fluctuations and the predictions from both models (Fig.~\ref{fig:WLC z(f)}(b)). However, due to the subtle cancellation effects, the total correction (Fig.~\ref{fig:WLC z(f)}(c)) is only correctly predicted by BRE-springs. 

Curiously, for WLC and BRE-springs the average of the force-elongation and the elongation-force relation for chains of finite length appears to be an excellent estimator for the asymptotic force-elongation curve (c/f panels (a) in Figs.~\ref{fig:BRE f(z)} and \ref{fig:BRE z(f)} as well as Figs.~\ref{fig:WLC f(z)} and \ref{fig:WLC z(f)} , or even more clearly the corresponding panels (c)). While this might be intuitively plausible, it is easy to show that the identity for BRE-springs is only approximate, but not exact.
Moreover, the examples of FENE- and BTB-springs would seem to indicate that this near identity is an accident rather than a rule (c/f Figs.~\ref{fig:FENE f(z)}(c) and \ref{fig:FENE z(f)}(c) as well as Figs.~\ref{fig:BTB f(z)}(c) and \ref{fig:BTB z(f)}(c)).
The corrections $\delta\zeta_{(\perp)}$ and $\delta\zeta_{(||)}$ due to transverse and longitudinal fluctuations only depend on the asymptotic force elongation relation, $\phi_{(\cdot)}(\zeta)$. For reasonable polymer models these corrections plausibly have a universal sign, because their origin is the relative stiffening of the springs on approaching their maximal elongation.
The different behavior of FENE-, BTB- and BRE-springs is due to the first order correction, $\delta\phi_{(\cdot)}(\zeta)$, which arises from the dominant contribution to the partition function from chain conformations with the minimal total elongation, $r=z$, at the considered projected elongation. While this terms vanishes for FENE-springs, for BTB- and BRE-springs it counteracts and largely cancels the fluctuation-induced corrections.

%
%
%

\begin{figure*}
\includegraphics[height=\textwidth, trim=35 0 0 0,clip]{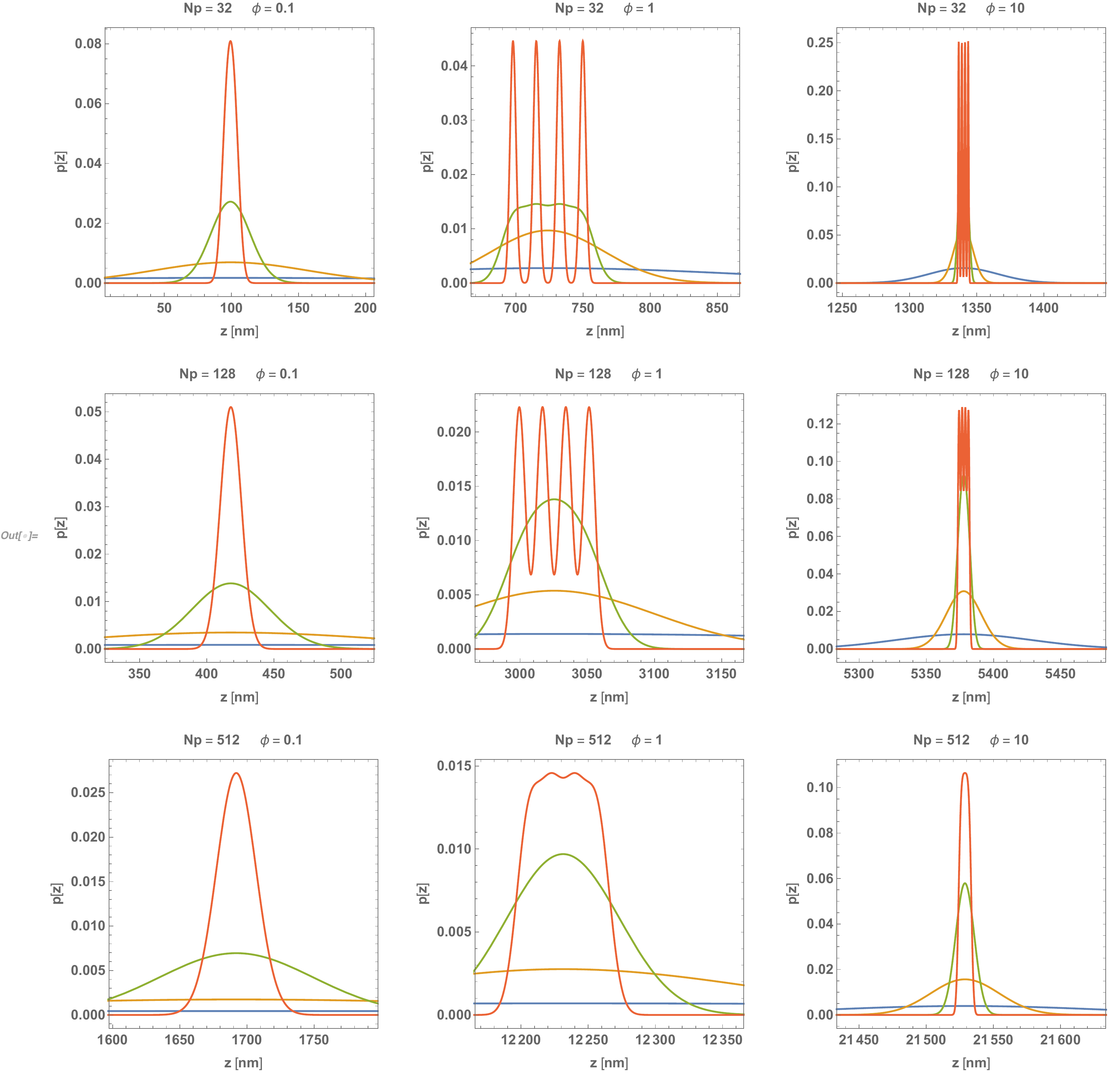}
\caption{
\label{fig:p(z) with Nicks}
Distribution of the parallel elongation for an ensemble of stretched ds-DNA molecules with $n-1=0,1,2,3$ nicks modelled as a corresponding sequence of freely-jointed BRE-springs.
The panels illustrate the effect of varying the total chain length, $N_p=32,128,512$, and the applied force, $\phi = f l_p/k_BT = 0.1,1.0,10.0$.
Blue: instantaneous elongations; Yellow, Green and Red: ``time averages'' over $N_{samples}=16,256,4096$ independent configurations.
Chain elongations are reported in ``nm'' for the example of ds-DNA with $l_p=50$nm. To simplify the comparison, all panels are centered on the elongation predicted by the asymptotic elongation-force relation, $L \zeta(\phi)$, and report $p(z)$ over $z$-values in intervals of an identical width of $4 l_p = 200$nm.
}
\end{figure*}

\subsection{Counting nicks in single-molecule stretching experiments of DNA}\label{sec:Nicks}
DNA single-molecule stretching experiments are typically performed on $\lambda$-phage DNA. Given the size of 48 kb or $N_p=320$ persistence lengths, finite-size effects are {\it a priori} of little concern. As a possible application of our results we discuss in the following the possibility to follow a dynamically changing number of  ``nicks''  in a molecule held at constant force by analysing the accompanying changes in the average elongation. Such a situation may arise in the presence of enzymes, which can induce and repair single-chain breaks.

Consider a defect-free ds-DNA segment of length $L$ under the influence of a dimensionless stretching force, $\phi$. Retaining finite chain length effects to first order, its average elongation is given by
\begin{eqnarray}\label{eq:quantized <z> for nicks}
\langle z \rangle
&\equiv& L \zeta(\phi) + l_p \delta_\zeta(\phi)\ ,
\end{eqnarray}
where $\delta_\zeta(\phi) = \lim_{\kappa\rightarrow0}  \delta \zeta(\phi,\kappa)/\kappa$.
This expression is straightforward to generalize to the situation, where the molecule is composed of $n$ {\em freely jointed} defect-free segments of a total length of $L=\sum_{i=1}^n  L_i$
\begin{eqnarray}
\langle z \rangle
&=& \sum_{i=1}^n \left( L_i \zeta(\phi) + l_p \delta_\zeta(\phi)\strut\right) \\
&=&  L \zeta(\phi) + n l_p \delta_\zeta(\phi)\ .
\end{eqnarray}
The above relation has a number of interesting implications:
(i) changing the number, $n-1$, of nicks by one changes the average chain elongation by a distance of the order of the DNA persistence length of $l_p=50$nm, and
(ii) this change neither depends on the total length, $L$, of the molecule nor on the precise position of the nicks.
The variance of the total elongation is again given by the sum of the variances of the subchain elongations. To zeroth order in $\kappa$, Eq.~(\ref{eq:k_|| BRE}) implies that
\begin{eqnarray}\label{eq: delta z^2}
\langle \delta z^2 \rangle
&=& \frac{l_p}{\phi'(\zeta^\ast(\phi))} \sum_{i=1}^n L_i = \frac{l_p L}{\phi'(\zeta^\ast(\phi))} \ .
\end{eqnarray}
With the standard deviation, $\sqrt{\langle \delta z^2 \rangle} \sim l_p \sqrt{N_p}$, increasing with chain length, the ``quantization'' is in general not observable in instantaneous configurations and emerges only in averages over time intervals, $T$, of sufficient length, where the relevant measure is the sampled number, $N_{samples}\sim T/\tau_{cor}$, of statistically {\em independent} configurations.
The correlation time, $\tau_{cor}$, for the fluctuating chain extension depends on the DNA dynamics in the experimental setup.
A simple blob picture~\cite{deGennesBook} would suggest that $\tau_{cor}\sim \phi^{-4}$ is a rapidly decreasing function of the applied force.

Fig.~\ref{fig:p(z) with Nicks} illustrates the influence of the chain length, $N_p$, of the applied stretching force, $\phi$, and of $N_{samples}$ on the distribution of (time-averaged) chain elongations for an ensemble composed of equal numbers of chains with $n-1=0,\ldots,3$ nicks.
There is obviously little point in exploring the effect of nicks in the weak stretching limit (l.h. column of Fig.~\ref{fig:p(z) with Nicks}).
While averaging over more and more statistically independent configurations sharpens the distributions around the mean, the peak does not split into separate peaks for molecules with different numbers of nicks.
This is easy to understand.
Firstly $\delta_\zeta(\phi)$ (and hence the distance between the quantized mean positions, Eq.~(\ref{eq:quantized <z> for nicks})) vanish in this limit (Figs.~\ref{fig:FENE z(f)}(c), \ref{fig:BRE z(f)}(c) and~\ref{fig:WLC z(f)}(c)).
Secondly, the chain fluctuations, Eq.~(\ref{eq: delta z^2}), are largest, because the effective spring constant for longitudinal fluctuations is a monotonously increasing function of the applied force.
The signal-to-noise ratio is better in the strong stretching limit (r.h. column of Fig.~\ref{fig:p(z) with Nicks}), but experiments might be challenging, since the {\em absolute} differences between the quantized elongations vanish again with $\delta_\zeta(\phi)$.
From an experimental point of view, the optimal regime is thus probably located around intermediate forces, $\phi\approx1$, where chains are stretched to about half their full elongation (central column of Fig.~\ref{fig:p(z) with Nicks}).
Comparisons between the three rows of Fig.~\ref{fig:p(z) with Nicks} illustrate the effect of chain length on the detection of nicks.
The effect of fluctuations decreases with $N_p$, if one considers the {\em relative} chain elongation, $\zeta=z/L$, which we have privileged throughout most of the article. However, in absolute terms, Eq.~(\ref{eq: delta z^2}), the width of the fluctuations {\em increases} with chain length.
Since the distances between the mean positions are independent of length (Eq.~(\ref{eq:quantized <z> for nicks})), $N_{samples}\sim N_p$ statistically independent configurations are expected to be needed to discriminate the number of nicks in the molecule.

\section{Summary and Conclusions}\label{sec:Conclusions}
The present work discusses the force-elongation and elongation-force relations of long polymer chains in single-molecule stretching experiments in the constant-force and in the constant-elongation ensemble. 
In particular, we show how to systematically derive these relations from a {\em given} radial end-to-end distance distribution and provide insight into the form and origin of the leading finite chain length corrections.   
The exactly solvable, non-trivial case of FENE-springs serves as a useful validation of our formalism and the employed numerical techniques.

In particular, we have used our formalism to explore the properties of ``BTB''- and ``BRE''-springs defined through approximate, closed analytical expressions for the end-to-end distance distribution of WLC~\cite{BTBpdf-1997,BeckerRosaEveraers2010}. 
While the BTB-distribution Eq.~(\ref{eq:QBTB}) derives from a variational treatment, the BRE-distribution Eq.~(\ref{eq:QBRE}) interpolates between all relevant, exactly known limiting cases from stiff to flexible chains and from looped to fully stretched configurations. For the present application to long WLC it was sufficient to analyse Eq.~(\ref{eq:QA}). 
To test the quality of the BTB- and BRE-approximations in the present context, we have performed Monte Carlo simulations of stretched WLC.

The asymptotic BRE force-elongation relation Eq.~(\ref{eq:QA fdot(z)}) reproduces the numerical solution of Marko's and Siggia's exact description~\cite{MarkoSiggia1995} to within $2$\%.
While probably less useful for experimental applications than the fit by Bouchiat {\it et al.}~\cite{BouchiatCroquetteBJ1999} of the exact MS relation, our formula has the merit of being the most precise among those resulting from a systematic theoretical approach to the problem~\cite{Vologodskii1994,MarkoSiggia1995,BouchiatCroquetteBJ1999}.

From our comparison to numerical data for WLC we tentatively conclude that the BRE-expressions for the finite chain lengths corrections in the two ensembles are of comparable quality. We argue that this precision might allow for an experimental application in the counting of ``knicks'' in single-molecule stretching experiments of ds-DNA, because their primary effect is the reduction of the effective chain length. 
As details on the form of the surface anchoring can lead to corrections of similar magnitude~\cite{KulicSchiesselPRE2005}, it might be difficult to count their {\em absolute} number. But the quantization of the mean elongations should allow to follow dynamic {\em changes} in the number of kinks provided they occur sufficiently slowly.

With respect to the theory of WLC, we notice that the partition function $\mathcal Z (f)$, Eq.~(\ref{eq:Partition function under force}), is equivalent to the Laplace-Fourier transform of the end-to-end distribution function $Q(\vec r)$ for which suitable sophisticate approximation schemes (such as the continued-fraction expansion of Ref.~\cite{SpakowitzPRE2008} or the Mathieu functions expansion for $2d$ WLC's of Ref.~\cite{KurzthalerFranoschPRE2017}) have been proposed. In future work, it might be interesting to explore, if these formalisms provide an alternative access to the asymptotic WLC force-elongation relation and to the finite-size corrections in the different ensembles.  

Finally, we speculate that the convenient mathematical properties of the FENE-model and our present results might be useful for the analysis~\cite{RiefPRL1998} of analogous experiments on protein and polysaccharides stretching, where the use of the WLC model is less pertinent than for ds-DNA.

\section*{Acknowledgements}
RE gratefully acknowledges discussions with G.S. Grest in a different context, which nevertheless triggered the present investigation.
Our work was supported by a STSM Grant from COST Action CA17139 (EUTOPIA). 
Furthermore we benefitted from stimulating discussions with D. Thirumalai during the ``Biological Physics of Chromosomes'' program organized at the Kavli Institute for
Theoretical Physics (Santa Barbara, USA) supported by NSF Grant No. PHY-1748958, NIH Grant No. R25GM067110, and the Gordon and Betty Moore Foundation Grant No. 2919.02. 
Finally we acknowledge the computer facilities of the FLMSN, notably of the P\^ole Scientifique de Mod\'elisation Num\'erique (PSMN) and the Centre Blaise Pascal (CBP) at the Ecole Normale Sup\'erieure de Lyon where simulations were performed. 

\appendix*

\section{Monte Carlo Simulations and Data Analysis}\label{sec:MCsimuls}
In the context of the Theory Section~\ref{sec:Theory} it was natural to first explore the constant-elongation ensemble and, in a second step, to use the obtained results as a basis for deriving the behavior in the constant-force ensemble.
For our numerical work it turns out to be easier to proceed in the opposite direction.
Section~\ref{sec:spring simulation Methods}
outlines (almost trivial) Monte Carlo simulations of stretched FENE-, BTB- and BRE-springs, while
Sec.~\ref{sec:WLC simulation Methods} 
briefly describes high-precision Monte Carlo (MC) computer simulations of a standard~\cite{BeckerRosaEveraers2010} numerical model of corresponding WLC.
In a second step, discussed in Sec.~\ref{sec:zstar Methods},
we calculate the {\it average} force at given {\it constant} elongation by analyzing the distribution function of spatial elongations in the constant-force ensemble.

\setcounter{subsection}{0}
\renewcommand\thesubsection{A.\arabic{subsection}}

\subsection{Elongation-force relations from Monte Carlo simulations of stretched FENE-, BTB- and BRE-springs}\label{sec:spring simulation Methods}
Given an analytic expression for the end-to-end distance distribution, $Q(r)$, the expectation value
\begin{equation}\label{eq:z(f) of Q}
\langle z(f) \rangle = \frac{\int d\vec r \,  Q(r)\,z \exp\left(\frac{f\, z}{k_BT}\right)}{\int d\vec r \,  Q(r)\exp\left(\frac{f\, z}{k_BT}\right) }
\end{equation}
is straightforward to sample using Metropolis Monte Carlo simulations~\cite{Metropolis1953}.
Starting from an arbitrary initial elongation, $\vec r$, with $|\vec r|\le L$, random changes of the end-to-end vector are accepted with a probability
\begin{equation}\label{eq:AccRatio-Q(r)}
\mathrm{acc}\left( \vec r \rightarrow \vec r\,' \right)
= \min\left( 1, \frac{Q(|\vec r\,'|) \exp\left(\frac{f\, z'}{k_BT}\right)}{Q(|\vec r|)\exp\left(\frac{f\, z}{k_BT}\right)}   \right)
\end{equation}
where $Q(|\vec r|>L) \equiv 0$.

Specifically, a single Monte Carlo step consists in the following.
At each given force $f$, we extract two uniformly distributed random numbers $|\vec r\,'| / L \in [0,1]$ and $z'/L \in [-1,+1]$ and move to this new position according to the probability Eq.~(\ref{eq:AccRatio-Q(r)}).
New positions are sampled each $10^3$ Monte Carlo steps, for a total of $10^6$ sampled positions per each force $f$ which corresponds to the statistics used for the WLC model (see Sec.~\ref{sec:WLC simulation Methods}).

Results for
FENE- ($Q(\vec r)$, Eq.~(\ref{eq:QFENE})),
BTB- ($Q(\vec r)$, Eq.~(\ref{eq:QBTB})),
and
BRE-springs ($Q(\vec r)$, Eq.~(\ref{eq:QBRE})) modeling polymer chains made of $N_p=4,8,16,32$ persistence lengths are shown in
Figs.~\ref{fig:FENE z(f)}, \ref{fig:BTB z(f)} and~\ref{fig:BRE z(f)} (symbols)
and are in excellent agreement with theoretical results (lines, see Secs.~\ref{sec:FENE}, \ref{sec:BTB} and~\ref{sec:BRE} for details). 
Reported error bars are calculated as the standard deviations of the corresponding means.

\subsection{Elongation-force relations from Monte Carlo simulations of moderately stretched WLC}\label{sec:WLC simulation Methods}

Results for WLC's of numerical quality comparable to the ones established for FENE-, BTB- and BRE-springs can be obtained from high-precision Monte Carlo (MC) computer simulations of the following standard~\cite{BeckerRosaEveraers2010} numerical model. 

We have considered linear polymer chains made of $N_b \equiv L/b = 512$ rigid bonds where 
$b$ is the bond length.
The energy of the chain is expressed 
by the Hamiltonian
\begin{equation}\label{eq:MCHam-ElongForce}
{\mathcal H} = {\mathcal H}_{\rm stiff} + {\mathcal H}_{\rm force} \, .
\end{equation}
${\mathcal H}_{\rm stiff}$ models the stiffness of the fiber and is given by
\begin{equation}\label{eq:Hstiff}
{\mathcal H}_{\rm stiff} = -k_{\rm stiff} \sum_{i=1}^{N_b-1} {\hat t}_i \cdot {\hat t}_{i+1}
\end{equation}
where ${\hat t}_i = \frac{{\vec r}_i - {\vec r}_{i-1}}b$ is the $i$-th unit bond vector and ${\vec r}_i$ ($i = 0, ..., N$) is the spatial position of the $i$-th bead.
The stiffness parameter $k_{\rm stiff}$ determines the persistence length $l_p$ of the polymer chain.
In fact, the bond-bond correlation function for ${\mathcal H}_{\rm force} = 0$ is given by
\begin{equation}\label{eq:TgTgCorrel}
\langle {\hat t}_{i+j} \cdot {\hat t}_i \rangle = \exp( - b |j| / l_p ) \, ,
\end{equation}
with
\begin{equation}\label{eq:lpKstiff}
\frac{l_p}{b} = - \frac1{\log \left( \coth \left( \frac{k_{\rm stiff}}{k_BT} \right) - \frac{k_BT}{k_{\rm stiff}} \right) } \ .
\end{equation}
It is easy to see that $\lim_{k_{\rm stiff} \rightarrow \infty} l_p / b \simeq k_{\rm stiff} / k_B T$.
The force term
\begin{equation}\label{eq:HbiasEF}
{\mathcal H}_{\rm force} = - \vec f \cdot (\vec r_{N_b} - \vec r_0) = - f (z_{N_b} - z_0)
\end{equation}
stretches the chain along the $z$-direction.

MC moves are based on the pivot algorithm~\cite{Sokal1994ArXiv,Sokal1996MC}.
A monomer $i$ between $0$ and $N_b-1$ is randomly selected and the portion of the chain comprising monomers $i$, ..., $N_b$ is rotated by an angle randomly picked in $[0,2\pi]$ around an axis centered on monomer $i$ and randomly oriented on the unit sphere. 
The move changes the set $\{\vec r \}$ of chain coordinates into $\{\vec r \,' \}$ and is then accepted according to the probability:
\begin{equation}\label{eq:AccRatioMetropolis}
\mathrm{acc}\left( \vec r \rightarrow \vec r\,' \right)
= \min\left( 1, \exp \left( - \frac{\mathcal H(\{\vec r \, '\}) - \mathcal H(\{\vec r \}) }{k_B T}\right) \right) \, .
\end{equation}
Single chain conformations are sampled at each $10^3N_b$ MC moves,
for a total of $10^6$ conformations per each force $f$.

As in the case of FENE-, BTB- and BRE-springs (Sec.~\ref{sec:spring simulation Methods}), we have simulated chains of a total length of $N_p = L/l_p = 32, 16, 8, 4$ persistence lengths, in order to be able to extrapolate to the asymptotic limit and to explore finite chain-length effects.
While ideally we would like to study WLC in the continuum limit with $b\rightarrow0$, we have obtained data for discrete bond lengths of $b/l_p=1/16,1/32,1/64,1/128 \ll 1$ or $N_b = 16, 32, 64,128 N_p$.
In particular, the choice of a bond length limits the range of forces, $f\ll k_BT/b$, which we can explore without encountering discretization effects.
In practical terms, we have sampled $f$ in the interval $[f_{\rm max}/1024, f_{\rm max}]$ in log-steps of $2$ with $f_{\rm max} = \frac1{16} \frac{k_BT}b$.
Corresponding results shown in Fig.~\ref{fig:WLC z(f)} (symbols) are in good agreement with theoretical results for BRE-springs (lines).

\subsection{Force-elongation relations from data obtained in the constant-force ensemble}\label{sec:zstar Methods}
The elongation-force relation in the constant-force ensemble is given by the sampled  average chain elongations, $\langle z(f) \rangle$.
In addition, one can sample corresponding histograms, $p_f(z)$.
Following the discussion in Sec.~\ref{sec:ElongationForce}, these histograms are peaked at an elongation $z^\ast(f)$, which is in general different from $\langle z(f) \rangle$. 
By correcting for the sampling bias due to the applied force, these histograms also 
provide a local estimate of the partition function, $\mathcal{Z}(z)$, in the constant elongation ensemble:
\begin{equation}
\mathcal{Z}(z) \propto p_f(z)\, \exp\left(-\frac{f\, z}{k_BT}\right)\ .
\end{equation}
This estimate will be efficiently sampled in the vicinity of $z^\ast(f)$, and multiple such local estimates could be tiled to estimate all of $\mathcal Z(z)$. 
Using Eq.~(\ref{eq:f(z)}), we can directly estimate the force-elongation relation over the sampled $z$-range as
\begin{equation}
\langle f(z) \rangle = f - k_BT\, \frac{ p_f'(z)} { p_f(z)} \ .
\end{equation}
In particular,
\begin{equation}
\langle f(z^\ast) \rangle = f
\end{equation}
at the peak of the sampled distribution, where the statistical quality of the results is highest.

The average location of the peak as a function of $f$ is determined as follows.
At each applied force $f$, we computed $10$ independent distributions $p_f(z)$ from the $10^6$ sampled elongations (Sec.~\ref{sec:spring simulation Methods}).
Then, the position of the peak of each distribution is estimated by the best fit of $\log p_f(z)$ to the function $a - \frac{k_2}2(z-z^\ast)^2 - {\rm sign}(z-z^\ast)\log(1+\frac{k_3}{6}|z-z^\ast|^3)$, {\it i.e.} the Gaussian function corrected for ``skewness'' with fit parameters $a$, $k_2$, $k_3$ and $z^\ast$.
We have found that the position of the maximum is accurately captured by limiting the fit to $\pm$ one standard deviation around the corresponding mean and estimating $p_f(z)$ from the histogram obtained by partitioning this interval into $40$ equally spaced bins.

The symbols in Figs.~\ref{fig:FENE f(z)}, \ref{fig:BTB f(z)}, \ref{fig:BRE f(z)} and \ref{fig:WLC f(z)} represent $\langle z^\ast(f)\rangle$ for FENE-springs, BTB-springs, BRE-springs and WLC respectively. Reported error bars indicate the standard error of the estimated means.

We take the excellent agreement of our numerical results for FENE-springs with the exact solution of the model (symbols {\it vs.} lines in Figs.~\ref{fig:FENE f(z)} and~\ref{fig:FENE z(f)}) as proof of the reliability of our method for converting between the two ensembles.
In the following, the numerical data for BRE-springs serve to validate our analysis of the asymptotic behavior of BRE-springs (Figs.~\ref{fig:BRE f(z)} and \ref{fig:BRE z(f)}) and can be directly compared to results for BTB-springs and WLC of finite length (Figs.~\ref{fig:WLC f(z)} and \ref{fig:WLC z(f)}).


\end{document}